\documentclass[acmsmall, screen, nonacm]{acmart}

\usepackage{mathrsfs}
\usepackage{mathpartir}
\usepackage{extarrows}
\usepackage{cancel}
\usepackage{makecell}
\usepackage{stmaryrd}
\usepackage{lineno}
\usepackage{subcaption}
\usepackage{tcolorbox}
\usepackage{enumitem}
\usepackage{changepage}
\usepackage{hyperref}

\usepackage{natbib}

\usepackage{amssymb}
\usepackage{tikz}
\usetikzlibrary{quantikz2}

\newcommand{\bC}{\mathbb{C}}

\newcommand{\bZ}{\mathbb{Z}}

\newcommand{\cD}{\mathcal{D}}
\newcommand{\cE}{\mathcal{E}}

\newcommand{\cP}{\mathcal{P}}

\newcommand{\cM}{\mathcal{M}}

\newcommand{\cH}{\mathcal{H}}

\newcommand{\cI}{\mathcal{I}}

\newcommand{\Tr}{\mathrm{Tr}}
\newcommand{\dom}{\mathrm{dom}}
\newcommand{\Dom}{\mathsf{Dom}}

\newcommand{\QHeap}{\mathsf{QHeap}}
\newcommand{\supp}{\mathrm{supp}}

\newcommand{\cnorm}[1]{{\left\vert\kern-0.25ex\left\vert\kern-0.25ex\left\vert #1 \right\vert\kern-0.25ex\right\vert\kern-0.25ex\right\vert}}
\renewcommand{\>}{\rangle}
\newcommand{\<}{\langle}
\renewcommand{\bar}[1]{\overline{#1}}
\newcommand{\pare}[1]{\left ( #1 \right )}

\newcommand{\purple}[1]{{\color{purple} #1}}
\newcommand{\blue}[1]{{\color{blue} #1}}
\definecolor{darkgreen}{rgb}{0.0, 0.4, 0.0}

\newcommand{\blueb}[1]{\blue{\left \{ #1 \right \}}}
\newcommand{\purpleb}[1]{\purple{\left \{ #1 \right \}}}
\newcommand{\triple}[3]{\blue{\left \{#1\right \}}~#2~\blue{\left \{#3\right \}}}

\newcommand{\bzero}{\mathbf{0}}

\newcommand{\sem}[1]{\left \llbracket #1 \right \rrbracket}

\newcommand{\tif}{\mathbf{if}}
\newcommand{\tthen}{\mathbf{then}}
\newcommand{\telse}{\mathbf{else}}

\newcommand{\twhile}{\mathbf{while}}
\newcommand{\tdo}{\mathbf{do}}

\newcommand{\tend}{\mathbf{end}}
\newcommand{\tskip}{\mathbf{skip}}

\newcommand{\sepimp}{\mathrel{-\mkern-6mu*}}
\newcommand{\otimesimp}{\mathrel{-\mkern-3mu\otimes}}
\newcommand{\talloc}{\mathbf{alloc}}
\newcommand{\trelease}{\mathbf{release}}

\newcommand{\gateCNOT}{\mathrm{CNOT}}

\newcommand{\sasaki}{\rightsquigarrow}

\newcommand{\TT}{{\mathtt{T}}}
\newcommand{\FF}{{\mathtt{F}}}

\usepackage{cleveref}
\usepackage{amsthm}

\theoremstyle{plain} 

\newtheorem{theorem}{Theorem}[section]
\newtheorem{proposition}[theorem]{Proposition}

\newtheorem{definition}[theorem]{Definition}
\newtheorem{example}[theorem]{Example}
\newtheorem{lemma}[theorem]{Lemma}

\begin{document}

\title{
    An Interpretation of Bunched Logic for Reasoning about Heap-Manipulating Quantum Programs
}
 
\author{Bonan Su} 
\orcid{0009-0009-7279-0658}
\email{sbn24@mails.tsinghua.edu.cn}
\affiliation{
    \department{Department of Computer Science and Technology}
    \institution{Tsinghua University}
    \city{Beijing}
    \country{China}
}

\author{Li Zhou}
\orcid{0000-0002-9868-8477}
\affiliation{
    \institution{Institute of Software, Chinese Academy of Sciences}
    \city{Beijing}
    \country{China}
}
\email{zhouli@ios.ac.cn}

\author{Yuan Feng}
\orcid{0000-0002-3097-3896}
\affiliation{
    \institution{Tsinghua University}
    \city{Beijing}
    \country{China}
}
\email{yuan_feng@tsinghua.edu.cn}

\author{Mingsheng Ying}
\orcid{0000-0003-4847-702X}
\affiliation{%
  \institution{University of Technology Sydney}
  \city{Sydney}
  \country{Australia}
}
\email{mingsheng.ying@uts.edu.au}

\begin{abstract}
    We introduce heap manipulation into quantum programming languages to enable flexible quantum memory management, which in turn poses new challenges for reasoning about program correctness.
    To address these challenges, we develop a novel quantum interpretation of bunched logic by extending Birkhoff--von Neumann quantum logic to Hilbert spaces whose dimensions vary to accommodate the dynamic allocation and deallocation of variables on the quantum heap.
    Beyond the \emph{separating conjunction} \((*)\), we present, for the first time, a quantum interpretation of the \emph{separating implication} \((\sepimp)\) to support backward reasoning.
    This interpretation preserves the adjunction between (\(*\)) and (\(\sepimp\)), ensuring that both operators capture spatial properties and are precisely aligned with the semantics of quantum heap manipulations.
    Building on this foundation, we establish a quantum separation logic that supports local reasoning and is relatively complete.

\end{abstract}

\maketitle

\section{Introduction}\label{sec:introduction}
Essential for large-scale deployment of quantum computing, \emph{quantum programming languages} have seen rapid development and widespread attention~\cite{qiskit, cirq, qsharp}.
Similar to classical programming languages, which support \emph{memory management} as a crucial mechanism to optimize resource utilization, many quantum programming languages also support the dynamic allocation and deallocation of quantum variables~\cite{qiskit, qsharpmemory}.

Beyond allocating fresh qubits in ground states, \emph{dirty qubits}~\cite{dirtyqubit,haner2016factoring,qsharpmemory} exploit ancillary qubits in unknown initial states, including entangled ones, to enhance qubit reuse and resource efficiency.
This motivates quantum programming languages to support the dynamic allocation of quantum variables in unknown initial states, analogous to the classical practice of allocating memory cells with uninitialized contents, thereby further reducing resource consumption.
However, allocating qubits with unknown initial states and potential entanglement poses additional challenges for ensuring correctness and memory safety, and is more subtle to reason about than in the classical setting, naturally underscoring the need for the development of formal methods tailored to analyzing such heap-manipulating quantum programs.

As one of the most successful formal methods for reasoning about programs with heap manipulations, \emph{separation logic}~\cite{bias, reynolds} adopts \emph{bunched logic}~\cite{logicofbi} as its assertion language to specify heap properties, and quantum separation logic is a well-suited approach to address the above challenges and has been actively studied~\cite{qsep,qsep1,qsep2,rapunsl}.
Crucially, building on an underlying logic like first-order logic, bunched logic introduces two multiplicative connectives---the \emph{separating conjunction} \((*)\) and the \emph{separating implication} \((\sepimp)\)---to describe properties of disjoint resources.

To motivate our key insight, we first explain, at a high level, the significance and interpretation of the two connectives in classical bunched logic, thereby justifying their suitability for quantum programs.
We then identify the technical challenges that arise in the quantum setting but remain unaddressed by previous work, and discuss how the classical interpretation should be adapted accordingly.
Finally, we summarize our solutions and contributions.

\subsection{Significance of Classical Bunched Logic}

\subsubsection{Separating Conjunction \((*)\): Spatially Local Heap Descriptions}
One of the most fundamental distinctions between stack and heap memory is that stack memory is accessed through static variable names, whereas heap memory is accessed through dynamically computed addresses, which are evaluated at runtime and may therefore be aliased or shared, as suggested by the phrase ``\emph{shared} mutable data structure'' in the title of Reynolds' foundational work~\cite{reynolds}.
This distinction leads to different principles for describing the two forms of
memory. Stack assertions naturally admit a \emph{global} interpretation: a
primitive assertion about program variables, such as \(x=3\), is interpreted
over any stack containing \(x\), independently of the presence and values of
other variables. Such a global interpretation, however, is inadequate for
capturing the desired properties of heap manipulations.

Due to the presence of memory sharing, manipulating one part of the heap may affect another.
Therefore, in addition to functional correctness, a comprehensive specification of heap manipulations should also account for memory safety: it should characterize not only how the relevant memory cells are changed, but also that the remaining memory cells are left unaffected---a property not captured by the global perspective.

For example, suppose primitive assertions for heap memory are interpreted globally, as in \(x\hookrightarrow 3\)\footnote{The notation \(x\hookrightarrow 3\) follows the standard notation in separation logic~\cite{reynolds,bias} and will be discussed formally later.}, which claims that the heap contains a memory cell at address \(x\) storing the value \(3\).
After releasing the memory cell at address \(x\), however, the resulting heap
can be described at best by \({\tt true}\).
The Hoare-style specification \(\triple{x\hookrightarrow 3}{\trelease(x)}{{\tt true}}\) then captures essentially no information about the effect of heap deallocation, since its postcondition is trivial.

To address this issue, assertions for local pieces of heap memory are introduced, where \(x\mapsto 3\) asserts that the heap contains \emph{exactly} one memory cell at address \(x\) storing the value \(3\).
Such exact and restrictive assertions, however, come at a cost: the traditional conjunction \((\wedge)\) is no longer suitable for combining them, since \(x\mapsto 3\wedge y\mapsto 4\) degenerates to \({\tt false}\) whenever \(x,y\) are distinct.
Therefore, a new logical connective, the \emph{separating conjunction} \((*)\), is introduced to spatially combine local assertions. 
Formally, for a heap \(h:\mathsf{Addr}\rightharpoonup\mathsf{Val}\) modeled as a partial function from addresses to values, the assertion \(P*Q\) is interpreted as
\begin{equation}\label{eq:sep-conj}
h\models P*Q\iff \exists h_1,h_2,\,h=h_1\uplus h_2 \mbox{ and } h_1\models P \mbox{ and } h_2\models Q,
\end{equation}
where \(\uplus\) denotes the disjoint union of partial functions.

The combination of \((\mapsto)\) and \((*)\) leads to ``\emph{remarkably simple axioms}''~\cite{bias} and enables precise specifications of heap manipulations, with their effects confined to the relevant heap fragments.
For the preceding example, it can now be specified as
\(
    \triple{x\mapsto 3}{\trelease(x)}{{\tt emp}},
\)
where \({\tt emp}\) asserts that the heap is empty.
This specification precisely captures the local effect of deallocation, and can be embedded into larger specifications by means of \((*)\) and the frame rule in separation logic---a bonus of \((*)\) to support \emph{local reasoning}.

\subsubsection{Separating Implication \((\sepimp)\): Adjunction for Backward Reasoning}
As the right adjoint of \((*)\), the predicate \(P\sepimp Q\) is formally interpreted as
\begin{equation}\label{eq:sep-imp}
h\models P\sepimp Q\iff \forall h',\,h'\models P \to h\uplus h'\models Q.
\end{equation}
The connection between \((*)\) and \((\sepimp)\) is demonstrated by the following equivalence, analogous to the adjunction between classical logical connectives (\(\wedge\)) and (\(\to\)):
\begin{equation}\label{eq:sep-duality}
P*Q \models R \quad \iff \quad P \models Q \sepimp R,
\end{equation}
where \(\models\) denotes logical entailment between predicates.
The separating conjunction and implication together enable a complete and elegant specification of heap-accessing assignments with general postconditions:
\begin{equation}\label{eq:assignment}
\triple{x \mapsto - * (x \mapsto v \sepimp P)}{[x]:=v}{P}.
\end{equation}
Intuitively, the precondition first ``removes'' the address \(x\) with value \(v\) from the heap using the predicate \(x \mapsto v\) and separating implication (\(\sepimp\)), and then ``reintroduces'' it with an unspecified value via the separating conjunction (\(*\)) and the predicate \mbox{\(x \mapsto -\)}.
Such specifications with general postconditions facilitate the establishment of a backward reasoning system.

\subsection{Suitability and Challenges of Bunched Logic for Quantum Programs}
As reviewed above, the separating conjunction \((*)\) and separating implication \((\sepimp)\) appear naturally suited to describing quantum heaps: their states are indexed by quantum variables, and heap manipulations are restricted to the target variables.
However, there are two key distinctions from classical heaps.
First, spatial disjointness is \emph{not} equivalent to state disjointness in the quantum setting. Second, spatially disjoint operations may nevertheless affect one another due to entanglement.

More specifically, classical heaps can be decomposed into subheaps with
disjoint domains without loss of information. By contrast, a quantum heap is
modeled as a density operator, which may be \emph{entangled} across subsystems.
Decomposing such a state into separate reduced states would in general lose
information about the original global state, and an operation on one subsystem
may affect another even when their targets are spatially disjoint.

These distinctions reveal a fundamental mismatch in existing works on quantum separation logic~\cite{qsep,qsep1,qsep2,rapunsl}: separating conjunction \((*)\) and separating implication \((\sepimp)\) are intended to describe spatial properties of heaps, whereas their existing interpretations in Eqs.~\eqref{eq:sep-conj} and~\eqref{eq:sep-imp} are based on decomposing the heap itself. 
Although these interpretations coincide with the intended spatial reading in the classical setting, this coincidence breaks down in the quantum setting, leading to the following issues and technical challenges.

\paragraph{Incompatibility between allocation and local reasoning} 
Existing approaches interpret \((*)\) in the most natural quantum way, following the state decomposition in Eq.~\eqref{eq:sep-conj} to ensure that operations on one subsystem do not affect the other. 
Viewing the tensor product \((\otimes)\) as the quantum analogue of classical heap division \((\uplus)\), one obtains the following \emph{separation-in-states} interpretation of \(P * Q\):
\[
\rho\models P*Q\iff \exists \rho_1,\rho_2,\,\rho=\rho_1\otimes \rho_2 \mbox{ and } \rho_1\models P \mbox{ and } \rho_2\models Q,
\]
where \(\rho,\rho_1,\rho_2\) represent quantum heaps.
Under this interpretation, satisfaction of \(P * Q\) requires the components described by \(P\) and \(Q\) to admit a tensor-product decomposition and therefore be \emph{unentangled}. 
Consequently, additional inference rules are required to handle entanglement~\cite{rapunsl}.

Although simple and well suited to local reasoning, this interpretation fails to specify quantum allocation, since it rules out entanglement. 
In particular, the assertion \(P * (q \mapsto -)\) cannot characterize the quantum heap obtained by allocating a fresh qubit \(q\) in an unspecified state that may be entangled with a heap satisfying \(P\), even though such allocations are natural in
practice, for instance when borrowing dirty qubits~\cite{dirtyqubit}. 
However, \(P * (q \mapsto -)\) is precisely the assertion one would expect to use for such spatial extensions, in accordance with the very nature of \((*)\) as a connective for spatial separation and of \((\mapsto)\) as a primitive for local pieces. 
This failure reveals a mismatch between the intended spatial reading of \((*)\) and its existing separation-in-states interpretation, which conflates spatial disjointness with state disjointness and is therefore overly restrictive.

\paragraph{Absence of Separating Implication (\(\sepimp\))} 
Existing approaches focus primarily on the separating conjunction \((*)\), while largely overlooking the separating implication \((\sepimp)\), whose quantum counterpart is considerably more subtle. 
Although \((\sepimp)\) can always be defined as the residual of \((*)\), either following Eq.~\eqref{eq:sep-imp} or, more generally, via a Galois connection once an interpretation of \((*)\) is fixed, such a definition may lack an intuitive meaning, preserve at most one direction of the adjunction in Eq.~\eqref{eq:sep-duality}, and fail to validate preconditions such as those in Eq.~\eqref{eq:assignment} for quantum operations. 
As discussed above, the lack of a satisfactory interpretation of \((\sepimp)\) complicates the formulation of weakest preconditions and backward reasoning rules.

\subsection{Our Solutions and Contributions}
To address the above limitations, we propose a new interpretation of the separating conjunction \((*)\) and separating implication \((\sepimp)\), with Birkhoff--von Neumann quantum logic~\cite{quantumlogic} as the underlying logic.
In contrast to existing approaches, the interpretation is based on \emph{separation in domains} rather than separation in states, thereby capturing the intended spatial meaning of \((*)\) and allowing entanglement between the separated domains.

Within this framework, for a given domain \(\tau\) (to be formally defined later), each predicate formula \(\psi\) is interpreted as a projection operator \(\sem{\psi}(\tau)\) on (or equivalently, a closed subspace of) the Hilbert space \(\mathcal{H}_{\tau}\) associated with \(\tau\).
The interpretations of the separating conjunction \(\psi * \varphi\) and the separating implication \(\psi \sepimp \varphi\) are then obtained by analogy with their classical counterparts in Eqs.~\eqref{eq:sep-conj} and~\eqref{eq:sep-imp}; more precisely, quantifiers (\(\exists\)) and (\(\forall\)) are now interpreted as join (\(\bigvee\)) and (possibly infinite) meet (\(\bigwedge\)) in the complete lattices of projection operators on $\mathcal{H}_\tau$, respectively:
\[
\begin{aligned}
    & \sem{\psi * \varphi}(\tau)
      &&= &&\bigvee _{\tau = \tau_1 \uplus \tau_2}
        \sem{\psi}(\tau_1) \otimes \sem{\varphi}(\tau_2)
      && \text{(cf.~Eq.~\eqref{eq:sep-conj})}, \\
    & \sem{\psi \sepimp \varphi}(\tau)
      &&= &&\bigwedge _{\tau':\tau'\cap \tau=\emptyset}
        \sem{\psi}(\tau') \otimesimp \sem{\varphi}(\tau \uplus \tau')
      && \text{(cf.~Eq.~\eqref{eq:sep-imp})},
\end{aligned}
\]
where \(\uplus\) denotes the disjoint union of domains, and (\(\otimesimp\)) is the adjoint operator of (\(\otimes\)) with respect to projection operators, as will formally be defined in Section~\ref{sec:bilogic}.

The interpretation of \(\psi * \varphi\) over a domain \(\tau\) proceeds by first decomposing \(\tau\) into two disjoint sub-domains \(\tau_1\) and \(\tau_2\), then combining the projections associated with \(\psi\) and \(\varphi\) on \(\tau_1\) and \(\tau_2\) via the tensor product \((\otimes)\), and finally taking the join \((\bigvee)\) over all such decompositions to reflect the existential quantification in Eq.~\eqref{eq:sep-conj}.
The interpretation of \(\psi \sepimp \varphi\) admits a similar but adjoint structure, which guarantees that the adjunction between (\(*\)) and (\(\sepimp\)) in Eq.~\eqref{eq:sep-duality} is fully preserved.
Moreover, as will be justified in Sections~\ref{sec:bilogic} and~\ref{sec:separation}, (\(*\)) and (\(\sepimp\)) are well suited for characterizing the allocation of quantum variables with unknown initial states and for expressing backward-reasoning rules, respectively.

As an application, we introduce a quantum programming language that supports dynamic allocation and deallocation of quantum variables in unknown initial states, and use the proposed quantum bunched logic as the assertion logic to develop a quantum separation logic for reasoning about functional correctness and memory safety of such heap-manipulating quantum programs. 

In particular, we obtain the following main results that have not been achieved in the previous research on quantum separation logic~\cite{qsep,qsep1,qsep2,rapunsl}: (1) our novel interpretation of (\(\sepimp\)) allows the logic to accommodate two schemes of inference rules: \emph{small axioms}~\cite{separationlogic} and backward reasoning; and (2) our quantum separation logic supports reasoning about both functional correctness and memory safety, and is relatively complete.

To keep the theory and presentation concise, we first restrict our attention to a purely quantum setting. 
In this setting, although disjointness of quantum variables can be determined statically at the syntactic level, it already suffices to expose the main technical challenges arising from genuinely quantum features and to demonstrate our solutions to them.
To account for pointer aliasing and memory sharing, which are crucial for practical programming, we further show in Section~\ref{sec:classical} that the framework can be naturally extended with classical variables and quantum arrays, thereby supporting reasoning about such phenomena in a hybrid classical--quantum setting.
Finally, we present case studies demonstrating the practical applicability of our logic to the verification of quantum programs involving the allocation of dirty qubits with unknown initial states, as well as while-loops annotated with invariants.



\section{Notations and Preliminaries}\label{sec:preliminary}

In this section, some notational conventions are introduced, without delving into the details or standard notations of quantum computing.
For a more comprehensive introduction, one may refer to standard textbooks~\cite{qcqi} on the subject.

\paragraph{Quantum Program States with Variables} Given a Hilbert space \(\cH\), which is assumed to be finite-dimensional throughout this paper, we use \(\cD(\cH)\) to denote the set of (partial) density operators on \(\cH\), i.e., positive semidefinite operators with trace less than or equal to 1. 
A \emph{quantum program state} is such an operator \(\rho\in \cD(\cH)\), whose trace \(\Tr(\rho)\) denotes the probability of the program being in state \(\rho\) during the execution of quantum programs.

For a quantum variable $q \in \mathsf{Var} = \{q_1, q_2, q_3, \ldots\}$, we write $\cH_q$ for its associated state space, and $\cH_{\bar{q}} = \bigotimes_{q \in \bar{q}} \cH_q$ for the joint state space of any finite set of quantum variables $\bar{q} \subseteq \mathsf{Var}$.
The dimension of \(\cH_q\) represents the type of the quantum variable \(q\), e.g., \(\cH_q\cong \bC^2\) suggests that \(q\) is a two-dimensional quantum variable (a qubit). 
In this paper, no additional notation will be introduced to track dimensions.
For example, the identity operator on \(\cH_q\) will simply be written as \(I\) without specifying its dimension, which is implicitly understood to match that of \(\cH_q\).

Similarly, we introduce subscripts to Dirac notation of density operators to indicate the correspondence between states and variables. 
For instance,
\(
\rho_{q_1,q_2}
  = |01\>_{q_1,q_2}\<01|
  = |10\>_{q_2,q_1}\<10|
\) denotes that the joint state of variables \(q_1\) and \(q_2\) is \(|01\>\<01|\).

Subscripts in unitary transformations \(U_{\bar{q}}\rho U_{\bar{q}}^\dagger\) and projective measurements \(P_{\bar{q}}\rho P_{\bar{q}}\) indicate the variables on which these operations act, and their dimensions are understood to match those of the corresponding unitary or projection operators. Identity operators on all other variables are implicitly omitted.
Moreover, we use the partial trace \(\Tr_{\bar{q}}(\rho)\) to denote the state obtained by tracing out the variables in \(\bar{q}\) from \(\rho\), and write \(\rho|_{\bar{q}}\) for the restriction of \(\rho\) to the variables in \(\bar{q}\), obtained by tracing out all other variables.

\paragraph{Projection Operators as Predicates and Birkhoff--von Neumann Quantum Logic} 
Given a Hilbert space \(\cH\), we use \(\cP(\cH)\) to denote the set of projection operators on \(\cH\), i.e., Hermitian operators \(P\) satisfying \(P^2=P\).
Each projection operator \(P\in \cP(\cH)\) naturally corresponds to a closed linear subspace of \(\cH\), which is also denoted by \(P\) with slight abuse of notation: \(P=\{|\psi\>\in \cH:P|\psi\>=|\psi\>\}\). 
As a set of pure states, a projection operator \(P\) can be viewed as a quantum predicate. 
A state \(\rho \in \mathcal{D}(\mathcal{H})\) satisfies \(P\), written \(\rho \models P\), if and only if the support space of \(\rho\) is contained within the subspace \(P\), i.e., \(\supp(\rho) \subseteq P\).
Here \(\supp(\rho)\) denotes the subspace spanned by the eigenvectors of \(\rho\) with non-zero eigenvalues.
Then, straightforwardly, the entailment between two predicates \(P_1, P_2 \in \cP(\cH)\), denoted \(P_1 \models P_2\), holds if and only if \(P_1 \subseteq P_2\) as subspaces.

By \emph{Birkhoff--von Neumann quantum logic}~\cite{quantumlogic} (or simply \emph{quantum logic}), 
we refer to the logic in which predicates are projection operators, and the logical 
connectives are defined following the orthomodular lattice structure of \(\cP(\cH)\):
\[
P_1 \wedge P_2 = P_1 \cap P_2, \quad 
P_1 \vee P_2 = \mathrm{span}(P_1 \cup P_2), \quad 
\neg P = P^\perp,
\]
where we make no notational distinction between a projection operator and its 
corresponding subspace, and \(P^\perp\) denotes the orthogonal complement of \(P\).
In addition, we introduce the \emph{Sasaki conjunction} (\(\Cap\)) and 
\emph{Sasaki implication} (\(\sasaki\))~\cite{sasaki}, defined by
\[
P \Cap Q = P \wedge (P^\perp \vee Q), \quad 
P \sasaki Q = P^\perp \vee (P \wedge Q),
\]
which together form a compatible pair of conjunction and implication 
analogous to classical logic. They satisfy the import-export condition
\(
P \Cap Q \models R \iff Q \models P \sasaki R,
\)
and \(P \models Q \iff P \sasaki Q = I\), where \(I\) denotes the identity operator.

These logical connectives are all established results in existing works and are not the focus of this paper. We introduce them briefly only to better align with classical logic, while our main focus is on the interpretation of multiplicative connectives in bunched logic.

\section{The Model of Quantum Heap}\label{sec:qheap}
Before introducing the logic, we first establish an intuitive yet formal model of the resource to be reasoned about, namely the \emph{quantum heap}.
We provide---through precise definitions and examples---a concise description of how the quantum heap is represented and manipulated, demonstrating its accommodation of many realistic scenarios.

As discussed in Section~\ref{sec:preliminary}, we use subscripts to record the correspondence between quantum states and variables.
Accordingly, we say that a quantum variable \emph{points to} a quantum state when the state is associated with the variable through its subscript.
For example, in the state \(\rho_{q_1,q_2} = |01\>_{q_1,q_2}\<01|\), the variable \(q_1\) points to the first qudit in state \(|0\>\), and \(q_2\) points to the second qudit in state \(|1\>\).

Similar to classical programming languages, we use statements of the form \(q := \talloc(d)\) to allocate a \(d\)-dimensional qudit and then let the variable \(q\) point to it.
However, as with classical pointers, repeated allocations may cause aliasing issues if the same variable is reused, potentially resulting in \emph{memory leaks}, where some allocated qudits become unreachable as no variable points to them. 
To model such scenarios, we introduce the symbol ``\(\sqcup\)'' in the subscripts, denoting qudits that have been allocated but are no longer pointed to by any variable. 

For example, starting from the state \(\rho_{q_1,q_2} = |01\>_{q_1,q_2}\<01|\) and performing a new qubit allocation via \(q_1 := \talloc(2)\), the resulting state may be \(|010\>_{\sqcup,q_2,q_1}\<010|\). 
Here, the first qudit becomes unreachable because \(q_1\) has been reassigned, while the third qubit is newly allocated in the quantum heap and now pointed to by \(q_1\).
Subscripts may include multiple \(\sqcup\) symbols, which serve solely as placeholders; the space dimension corresponding to each \(\sqcup\) is not explicitly tracked.

Now we are ready to present the formal definition of quantum heaps.
\begin{definition}[Quantum Heap]\label{def:quantum-heap}
    The set of quantum heaps is defined as partial density operators with finitely many variables in the subscripts, where some of these variables are replaced by the special symbol ``\(\sqcup\)''.
    That is,
    \[
    \QHeap\triangleq\bigcup_{\bar{q}\subseteq_{\mathrm{fin}}\mathsf{Var}}\bar{\cD}(\cH_{\bar{q}}),
    \]
    where \(\bar{\cD}(\cH_{\bar{q}})\) denotes the set of partial density operators on \(\cH_{\bar{q}}\) with some subscripts replaced by ``\(\sqcup\)''. 
\end{definition}
In particular, a quantum heap over a one-dimensional space degenerates to a real number in the interval \([0,1]\), representing the probability mass of the heap in the absence of any state information.
It is worth distinguishing between \emph{empty heaps} and \emph{zero heaps}: the former are heaps defined over a one-dimensional space, whereas the latter are heaps whose trace is zero.

In addition, for any quantum heap \(\rho \in \QHeap\), we use \(\dom \rho\) to denote the collection of its subscripts, which can be viewed as the \emph{domain} of \(\rho\) from an operational perspective. 
We further write \(\Dom\) for the collection of all possible domains of quantum heaps, whose elements are unions of a finite set of variables and multiple ``\(\sqcup\)'' symbols.

As examples, the previously mentioned states 
\(
|01\>_{q_1,q_2}\<01|,\ |010\>_{\sqcup,q_2,q_1}\<010|\in\QHeap
\) 
are both quantum heaps.
Next, we examine the formal behaviors of the three fundamental heap operations: mutation, allocation, and deallocation.
These operations will be formalized as semantics of statements in a quantum programming language in later sections.

\paragraph{Mutation of Quantum Heaps} 
Following the classical programming convention, given a quantum operation, we enclose the variables in square brackets to denote the target qudits it acts upon. 
Taking unitary transformation as an example, the statement \(\gateCNOT[q_2,q_1]\) applies the controlled-NOT gate to the qudits pointed to by variables \(q_2\) and \(q_1\).
The state obtained by applying this statement to the quantum heap \(|01\>_{q_1,q_2}\<01|\) is thus \(|11\>_{q_1,q_2}\<11|\). 
For simplicity, we omit checks for qudit-dimension consistency, which are analogous to type checking in classical settings.

Notably, only variables in \(\mathsf{Var}\) can serve as targets of mutation operations, while the special symbol ``\(\sqcup\)'' cannot, ensuring that mutations are never applied to unreachable qudits. If some target variables are absent from the domain of the quantum heap, i.e., \(\bar{q} \not\subseteq
\dom\rho\), the execution of the mutation statement is considered \emph{stuck}, representing a runtime error that will be formalized in later discussions on formal semantics of quantum programs.

\paragraph{Allocation of New Qudits} 
When a new qudit is allocated via the statement \(q := \talloc(d)\), its execution on a quantum heap \(\rho\) proceeds naturally in two steps:
\begin{enumerate}
    \item First, every occurrence of \(q\) in the subscripts of \(\rho\) is replaced with ``\(\sqcup\)'', indicating that the corresponding qudits are no longer reachable due to the reassignment of variable \(q\).
    \item Next, a new \(d\)-dimensional qudit in an arbitrary state is nondeterministically appended to the quantum heap, and variable \(q\) is set to reference it.
\end{enumerate}
The second step requires further explanation, as it introduces nondeterminism to model the unknown initial state of the newly allocated qudit, analogous to memory allocation in classical programming languages.

By appending a qudit in an unknown state, the quantum heap \(\rho\) is transformed into a new heap \(\rho'\) satisfying \(\dom \rho' = \dom \rho \cup \{q\}\) and \(\rho'|_{\dom \rho} = \rho\), so that tracing out the newly added qudit from \(\rho'\) recovers the original heap \(\rho\).
This nondeterministic choice of \(\rho'\) extends the classical notion of allocating a memory cell with an unknown value to the quantum setting, as the only constraint is that the reduced state on the original domain remains unchanged.

For example, a possible execution of \(q_1 := \talloc(2)\) on the quantum heap \(|01\>_{q_1,q_2}\<01|\) is demonstrated as follows:
\[
\begin{aligned}
|01\>_{q_1,q_2}\<01|& \xrightarrow{\text{Step 1}} |01\>_{\sqcup,q_2}\<01|\xrightarrow{\text{Step 2}} |010\>_{\sqcup,q_2,q_1}\<010|.
\end{aligned}
\]
It is worth noting that this transformation provides considerable flexibility in the initial state of the newly allocated qudit, particularly for mixed states as illustrated in the following example.
\begin{example}
    Consider the quantum heap \(\frac{1}{2}|0\>_{q_1}\<0|+\frac{1}{2}|1\>_{q_1}\<1|\in \QHeap\).
    For the statement \(q_2 := \talloc(2)\), the following are three valid resulting heaps \(\rho'\):
    \begin{enumerate}
        \item \(\rho'=\bigl(\tfrac{1}{2}|0\>_{q_1}\<0|+\tfrac{1}{2}|1\>_{q_1}\<1|\bigr)\otimes |0\>_{q_2}\<0|\), 
        where the newly allocated qudit is isolated and in the pure state \(|0\>\);

        \item \(\rho'=\tfrac{1}{2}(|00\>+|11\>)_{q_1,q_2}(\<00|+\<11|)\), 
        where the original mixed state is a subsystem of an entangled state, and the allocation reintroduces the remaining part of that entangled system;

        \item \(\rho'=\bigl(\tfrac{1}{2}|0\>_{q_1}\<0|+\tfrac{1}{2}|1\>_{q_1}\<1|\bigr)\otimes 
        \bigl(\tfrac{1}{2}|0\>_{q_2}\<0|+\tfrac{1}{2}|1\>_{q_2}\<1|\bigr)\), 
        where the newly allocated qudit can belong to a different entangled system.
    \end{enumerate}
    All these resulting heaps satisfy the requirement that tracing out \(q_2\) yields the original heap, and they reflect different possible operational scenarios.
\end{example}

\paragraph{Deallocation of Qudits}
Similar to the mutation of quantum heaps, the deallocation of a qudit must also be performed through the variable that points to it.
Specifically, the statement \(\trelease(q)\) simply traces out the qudit pointed to by the variable \(q\) from the quantum heap.

For example, executing the release statement \(\trelease(q_2)\) on the quantum heap \(|01\>_{q_1,q_2}\<01|\) will result in the quantum heap \(|0\>_{q_1}\<0|\).
It is important to note that only variables in \(\mathsf{Var}\) can be released, whereas qudits indexed by ``\(\sqcup\)'' cannot. 
Moreover, for any quantum heap obtained through an allocation, releasing the newly allocated qudit always recovers the original heap prior to the allocation, regardless of the nondeterminism in its initial state.

\section{A Quantum Bunched Logic}\label{sec:bilogic}
In this section, we present a novel quantum interpretation of bunched logic.
After introducing its formal syntax and semantics, we establish a proof system for reasoning about predicate entailments.
\subsection{Syntax and Semantics}\label{sec:bilogic-semantics}
In the logic of bunched implications~\cite{logicofbi}, predicates are typically interpreted over formal semantic structures, most notably resource semantics in the form of BI frames and categorical semantics based on doubly closed categories.
However, we will not instantiate or reinterpret such frames or categories in the quantum setting, as done in previous work~\cite{qsep}.
Instead, we present a quantum interpretation of bunched logic by extending Birkhoff--von Neumann quantum logic with the separating conjunction \((*)\) and separating implication \((\sepimp)\), thereby enabling the specification of properties of quantum heaps.

A fundamental distinction between our logic and Birkhoff--von Neumann quantum logic lies in the underlying Hilbert space.
While the latter is defined over a fixed Hilbert space, our logic is interpreted over a \emph{variable} Hilbert space that dynamically expands or contracts as quantum variables are allocated or deallocated.
Compared with existing quantum interpretations of bunched logic, our approach is more closely aligned with standard quantum logics and more faithfully captures the inherently quantum nature of heap-based resources.

\begin{figure*}[t]
    \centering
    \begin{minipage}{0.45\linewidth}
        \centering
        \begin{tikzpicture}[remember picture, overlay]
            \fill[opacity=1.0, fill=blue!5, dashed, rounded corners=5pt] (-1.7,-1.65) rectangle (3.3,-0.1);
            \draw (2.05, -1.4) node[text=blue] {(Classical logic)};
            \fill[opacity=1.0, fill=purple!5, dashed, rounded corners=5pt] (-1.7,-2.6) rectangle (3.3,-1.7);
            \draw (1.25, -2.4) node[text=purple] {(Multiplicative connectives)};
        \end{tikzpicture}
        \[
        \begin{aligned}
        \psi::=\ &\top\mid\bot\mid p\in\mathsf{underlying\ logic}\\
        \mid\ &\psi_1\wedge\psi_2\mid\psi_1\vee\psi_2\mid\neg\psi\\
        \mid\ & \psi_1\rightarrow\psi_2\\
        \mid\ & \top^*\mid\psi_1*\psi_2\mid \psi_1\sepimp\psi_2\\~\\
        \end{aligned}
        \]
        \subcaption{Classical bunched logic formula}
    \end{minipage}
    \hspace{2em}
    \begin{minipage}{0.45\linewidth}
        \centering
        \begin{tikzpicture}[remember picture, overlay]
            \fill[opacity=1.0, fill=blue!5, dashed, rounded corners=5pt] (-2.4,-1.15) rectangle (3.3,-0.1);
            \draw (1.1, -0.9) node[text=blue] {(Birkhoff-von Neumann logic)};
            \fill[opacity=1.0, fill=orange!5, dashed, rounded corners=5pt] (-2.4,-1.65) rectangle (3.3,-1.2);
            \draw (2.7, -1.4) node[text=orange] {(Sasaki)};
            \fill[opacity=1.0, fill=purple!5, dashed, rounded corners=5pt] (-2.4,-2.6) rectangle (3.3,-1.7);
            \draw (1.25, -2.4) node[text=purple] {(Multiplicative connectives)};
        \end{tikzpicture}
        \[
        \begin{aligned}
            \psi::=\ & \top\mid\bot\mid \bar{q}\mapsto P\mid \neg\psi\mid\psi_1\vee\psi_2\mid \psi_1\wedge\psi_2\\
            ~\\
            \mid\ & \psi_1\Cap\psi_2\mid\psi_1\sasaki\psi_2\\
            \mid\ & \top^*\mid\psi_1*\psi_2\mid\psi_1\sepimp\psi_2\\
            ~\\
        \end{aligned}
        \]
        \subcaption{Quantum bunched logic formula}
        \label{fig:qbi-formula}
    \end{minipage}
    \caption{Comparison between classical bunched logic and quantum bunched logic}
    \label{fig:compare}
\end{figure*}

Figure~\ref{fig:compare} compares formulas in classical bunched logic with those in our quantum bunched logic. Since the focus is on the bunched structures, we omit the details of the underlying logics and use \(p\) to denote a predicate; for instance, \(p \equiv \exists y,\ x=2\cdot y\) in first-order arithmetic logic. 
Nevertheless, logical connectives (\(\wedge, \vee, \neg, \to\)) are explicitly included, motivated by the well-established analogies in Birkhoff--von Neumann quantum logic~\cite{quantumlogic} with Sasaki's implication~\cite{sasaki}, from which their interpretations in our logic are naturally inherited.

Next, we present the formal semantics of the quantum bunched logic, with particular emphasis on \((*)\) and \((\sepimp)\). 
Since the underlying logic is not the main focus, we do not follow the standard approach of introducing structures or models to interpret primitive symbols. 
Instead, we take the symbol \(P\) in \(\bar{q}\mapsto P\) simply as an arbitrary projection operator, treating the underlying logical formula directly via its semantics from an extensional perspective.

Given a domain \(\tau \in \Dom\) (a finite set containing variables in \(\mathsf{Var}\) together with possibly multiple occurrences of the symbol~\(\sqcup\), as introduced after Definition~\ref{def:quantum-heap}), each predicate \(\psi\) is interpreted inductively as a projection operator \(\sem{\psi}(\tau)\) on the Hilbert space associated with \(\tau\), as shown in Figure~\ref{fig:qbi-semantics}. 



With the idea of variable Hilbert spaces introduced earlier in this section, the interpretation of Birkhoff--von Neumann logic can be naturally extended to our quantum bunched logic. 
Throughout, we implicitly assume that the dimensions of all operators are compatible with their domains. For instance, in an assertion \(\bar{q} \mapsto P\), the operator \(P\) is a projection on \(\mathcal{H}_{\bar{q}}\), while \(\bzero\) and \(I\) denote the zero and identity operators on the corresponding Hilbert spaces.
Under this variable-domain interpretation of quantum bunched logic, the semantics of \(\sem{\psi_1 * \psi_2}(\tau)\) and \(\sem{\psi_1 \sepimp \psi_2}(\tau)\) coincide with the many-valued logical counterparts of Eqs.~\eqref{eq:sep-conj} and \eqref{eq:sep-imp}, respectively, as outlined in Section~\ref{sec:introduction}. 

\begin{figure}
    \centering
    \[
    \begin{aligned}
        &\sem{\top}(\tau) = I,\qquad \sem{\bot}(\tau) = \bzero,\qquad\sem{\bar{q}\mapsto P}(\tau)=\begin{cases}
            P, & \mbox{if }\tau=\bar{q},\\
            \bzero, & \mbox{otherwise},
        \end{cases}\qquad\sem{\neg \psi}(\tau) = \pare{\sem{\psi}(\tau)}^\perp,\\
        &\sem{\psi_1 \bowtie \psi_2}(\tau) =\sem{\psi_1}(\tau) \bowtie \sem{\psi_2}(\tau),\ \mbox{for }\bowtie\,\in\{\wedge,\vee,\Cap,\sasaki\}\qquad
        \sem{\top^*}(\tau) = \begin{cases}
            1, & \mbox{if }\tau=\emptyset,\\
            \bzero, & \mbox{otherwise},
        \end{cases}\\
        &\sem{\psi_1 * \psi_2}(\tau) = \bigvee_{\substack{\tau_1\cup \tau_2=\tau\\\tau_1\cap \tau_2=\emptyset}}\sem{\psi_1}(\tau_1) \otimes \sem{\psi_2}(\tau_2),\qquad
        \sem{\psi_1 \sepimp \psi_2}(\tau) = \bigwedge_{\substack{\tau'\in\Dom\\\tau'\cap \tau=\emptyset}}\sem{\psi_1}(\tau')\otimesimp\sem{\psi_2}(\tau'\cup\tau).
    \end{aligned}
    \]
    \caption{Semantics of quantum bunched logic, where \(P \otimesimp Q\) denotes the largest projection operator \(R\) such that \(P \otimes R \subseteq Q\), as will be defined in Lemma~\ref{lemma:tensor-implication}.}
    \label{fig:qbi-semantics}
\end{figure}

The satisfaction relation between a quantum heap \(\rho \in \QHeap\) and a predicate \(\psi\) is naturally defined by
\[
\rho \models \psi \;\iff\; \supp(\rho) \subseteq \sem{\psi}(\dom \rho),
\]
which requires that the support space of the quantum heap lies within the subspace determined by the interpretation of \(\psi\) over the domain of \(\rho\).
In particular, over empty domains, quantum heaps degenerate to real numbers in \([0,1]\) as discussed after Definition~\ref{def:quantum-heap}.
In this case, the semantics of predicates also collapse to Boolean values, namely \(0\) or \(1\). 
Consequently, when \(\dom \rho = \emptyset\), we have
\(
  \rho \models \psi\) if and only if \(\rho \leq \sem{\psi}(\emptyset)
\).
\begin{example}[Empty Heaps and Zero Heaps]
    For an empty heap \(\rho = \tfrac{1}{2}\), we have \(\rho \models \top^*\) since \(\sem{\top^*}(\emptyset)=1\), whereas \(\rho \not\models \bot\) because \(\sem{\bot}(\emptyset)=0\).
In contrast, for the empty and zero heap \(\rho' = 0\), we have both \(\rho' \models \top^*\) and \(\rho' \models \bot\), which further illustrates that the zero heap satisfies any predicate and is generally not considered.
\end{example}

Next, we delve into the meaning and specification of predicate interpretations, and provide several examples to illustrate the underlying intuition.



\paragraph{Assertion on Fragments of Heaps: \(\bar{q}\mapsto P\)}
Analogous to the predicate \(x \mapsto v\) in classical bunched logic, which asserts that the heap contains exactly a \emph{single} memory cell with address \(x\) and value \(v\), the predicate \(\bar{q} \mapsto P\) in our quantum bunched logic asserts that the quantum heap contains only the quantum variables in \(\bar{q}\), with the support space of their joint state contained in the subspace defined by the projection operator \(P\).

For simplicity, we may assume that the dimension of the operator \(P\) is compatible with the joint state of the variables in \(\bar{q}\). 
Note that if the domain of a quantum heap \(\rho\) does not coincide with \(\bar{q}\), then \(\rho\) cannot satisfy \(\bar{q} \mapsto P\), as \(\sem{\bar{q} \mapsto P}(\dom \rho) = \bzero\) represents the zero subspace which is only satisfied by the zero heap.
\begin{example}
    The quantum heap \(\rho=|01\>_{q_1,q_2}\<01|\) satisfies the predicate \(q_1,q_2\mapsto |01\>\<01|\) since 
    \[
    \sem{q_1,q_2\mapsto |01\>\<01|}(\{q_1,q_2\})=|01\>_{q_1,q_2}\<01|\supseteq \supp(\rho).
    \]
    However, once \(q_1\) is reallocated as discussed in the previous section, the resulting heap \(\rho'=|010\>_{\sqcup,q_2,q_1}\<010|\) will no longer satisfy the predicate as \(\sem{q_1,q_2\mapsto |01\>\<01|}(\{\sqcup,q_1,q_2\})=\bzero\).
\end{example}


\paragraph{The Separating Conjunction \((*)\)}
In classical bunched logic, the separating conjunction is interpreted by partitioning the heap into two disjoint subheaps, each of which satisfies one of the clauses as formally defined in Eq.~\eqref{eq:sep-conj}.

Analogously---though not in exactly the same manner---we interpret the separating conjunction in our quantum bunched logic by dividing the domain \(\tau\) into two disjoint parts \(\tau_1, \tau_2\) with \(\tau_1 \cap \tau_2 = \emptyset\) and \(\tau = \tau_1 \cup \tau_2\).
Each clause predicate is interpreted on one part of the domain, and the two resulting projectors are then combined via the tensor product, ensuring that the outcome remains a projection operator on the Hilbert space associated with \(\tau\).

Since the predicate is satisfied once such divisions exist, we take the disjunction (for projection operators) over them.
It is worth noting that disjunction in quantum logic is weaker than in classical logic, in the sense that \(\rho \models \psi * \varphi\) does not necessarily imply the existence of a specific partition \(\tau = \tau_1 \cup \tau_2\) such that 
\(
\supp(\rho) \subseteq \sem{\psi}(\tau_1) \otimes \sem{\varphi}(\tau_2),
\) 
which marks a departure from the classical case and is demonstrated in the following concrete example.
\begin{example}
    Consider the predicate \(\psi\equiv q_1\mapsto |0\>\<0|\vee q_1,q_2\mapsto |11\>\<11|\). By definition, we have 
    \[
    \begin{aligned}
    \sem{\psi*\top}(\{q_1,q_2\})&=\sem{\psi}(\{q_1\})\otimes I_{q_2}\vee \sem{\psi}(\{q_1,q_2\})\\
    &=|0\>_{q_1}\<0|\otimes I_{q_2}\vee |11\>_{q_1,q_2}\<11|\\
    &=|00\>_{q_1,q_2}\<00| + |01\>_{q_1,q_2}\<01| + |11\>_{q_1,q_2}\<11|.
    \end{aligned}
    \]
    Therefore, \(\rho=|\!+\!1\>_{q_1,q_2}\<+1|\) satisfies the predicate \(\psi * \top\), since 
    \[
    \supp(\rho)=\mathrm{span}(\{|\!+\!1\>\})\subseteq \mathrm{span}(\{|00\>,|01\>,|11\>\}).
    \]
    However, neither \(\supp(\rho)\subseteq \sem{\psi}(\{q_1\})\otimes \sem{\top}(\{q_2\})\), nor \(\supp(\rho)\subseteq \sem{\psi}(\{q_1,q_2\})\otimes \sem{\top}(\emptyset)\) holds, illustrating that no specific partition of the domain satisfies the two clauses individually.
\end{example}

Furthermore, in contrast to separation in heaps, the separation-in-domain scheme permits entanglement between the two parts of the domain, since the tensor product of two projection operators does not preclude entanglement in the combined space. 
The following example illustrates the interpretation of a specific predicate, emphasizing that entangled states can also satisfy predicates involving the separating conjunction.

\begin{example}
    The interpretation of the predicate \(\psi * q \mapsto I\) over the domain \(\tau \cup \{q\}\) is given by
\[
\sem{\psi * q \mapsto I}(\tau \cup \{q\}) = \sem{\psi}(\tau) \otimes I_q.
\]
Consequently, for any quantum heap \(\rho\) with \(\dom(\rho) = \tau \cup \{q\}\), if \(\Tr_q(\rho) \models \psi\), then \(\rho \models \psi * q \mapsto I\) by definition, regardless of whether \(\rho\) is entangled between the subsystem associated with \(\tau\) and the qudit \(q\).
Hence, the predicate \(\psi * q \mapsto I\) precisely characterizes the heap obtained by allocating a fresh qudit \(q\) in an arbitrary initial state to \(\Tr_q(\rho)\), as discussed in Section~\ref{sec:qheap}.
\end{example}

Despite the possible presence of entanglement, the interpretation suffices to ensure
that a local quantum operation on one part of the domain does not affect the other.
This property will be formalized as the \textsc{Frame} rule in Section~\ref{sec:separation}; here, we provide only an intuitive explanation.
For two projection operators \(P,Q\) and a quantum heap \(\rho\) with \(\supp(\rho) \subseteq P \otimes Q\), it can be verified that applying a quantum operation \(\cE\) on the domain of \(P\) produces a new heap \(\rho' = (\cE \otimes \cI)(\rho)\) satisfying \(\supp(\rho') \subseteq P' \otimes Q\) for some projection operator \(P'\), leaving the part \(Q\) unaffected.

Analogous to the notations in classical bunched logic~\cite{reynolds, bias}, we
introduce the syntactic sugar \(\bar{q} \hookrightarrow P \equiv \bar{q} \mapsto P * \top\) to indicate that the quantum heap contains at least the variables in \(\bar{q}\), and that the projection operator \(P\) is extended to the entire heap by tensoring with the identity on the remaining variables.
For example, it can be verified by definition that for a quantum heap \(\rho=|01\>_{q_1,q_2}\<01|\), \(\rho\models q_1\hookrightarrow |0\>\<0|\) but \(\rho\not\models q_1\mapsto |0\>\<0|\).

\paragraph{The Separating Implication \((\sepimp)\)}
There are two primary intuitions about separating implication, which is also referred to as the \textit{magic wand} in certain contexts: (1) \(\psi_1\sepimp\psi_2\) is the weakest predicate \(\varphi\) such that \(\psi_1*\varphi\) entails \(\psi_2\); and (2) \(\rho\vDash\psi_1\sepimp\psi_2\) suggests that, for any expansion \(\rho'\)  (not necessarily in the form of \(\rho\otimes \sigma\)) of \(\rho\) with \(\rho'|_{\dom\rho}=\rho\), if the expanded part satisfies \(\psi_1\), then the whole heap \(\rho'\) will satisfy \(\psi_2\) as formally defined in Eq.~\eqref{eq:sep-imp}.
The interpretation of the separating implication (\(\sepimp\)) in our logic precisely captures the second intuition, and we will show that it also satisfies the first intuition after introducing the entailment relation between predicates.

Specifically, in the interpretation of the separating implication (\(\sepimp\)), we introduce a novel operator, \emph{tensor implication} (\(\otimesimp\)), to formulate the adjoint operator of (\(\otimes\)) with respect to projection operators.
Intuitively, \(P \otimesimp Q\) denotes the largest projection operator \(R\) on the complementary domain such that \(P \otimes R \subseteq Q\); 
the existence of \(R = P \otimesimp Q\) is guaranteed by the following lemma.
\begin{lemma}[Tensor Implication]\label{lemma:tensor-implication}
    Let \(P\) and \(Q\) be two projections with \(\dom P \subseteq \dom Q\). Define
    \[
    P \otimesimp Q \triangleq
        \max \{ R \in \cP(\cH_{\dom Q \backslash \dom P}) : P \otimes R \subseteq Q \}.
    \]
    Then \(P \otimesimp Q = I\) if \(P = \mathbf{0}\), and otherwise
    \[
    P \otimesimp Q = E\Big(\Tr_{\dom P}\Big(\frac{P}{\dim P} \cdot Q\Big)\Big),
    \]
    where \(\dim P\) denotes the dimension of \(P\) and \(E(A)\) denotes the
    subspace spanned by the eigenvectors of \(A\) with eigenvalue 1.
    Moreover, for any \(R\), \(P \otimes R \subseteq Q\) if and only if
    \(R \subseteq P \otimesimp Q\), which establishes the adjunction between
    (\(\otimes\)) and (\(\otimesimp\)).
\end{lemma}
The interpretation of predicates involving separating implication in
Figure~\ref{fig:qbi-semantics} can be intuitively understood as follows.
For any disjoint domain \(\tau'\) (which can be viewed as the expanded part), the quantum heap
\(\rho\) satisfies \(\psi_2\) (over \(\tau \cup \tau'\)) ``excluding'' \(\psi_1\) (over \(\tau'\)).
This condition suffices to ensure that any extension of \(\rho\) to the domain \(\tau \cup \tau'\) satisfies \(\psi_2\), provided that the added part satisfies \(\psi_1\), since the conjunction (\(\bigwedge\)) corresponds to universal quantification over all disjoint extensions. 
The ``exclusion'' is implemented via (\(\otimesimp\)), which serves as the adjoint operator of (\(\otimes\)) with respect to projection operators, as discussed above. 
The combination of (\(\bigwedge\)) and (\(\otimesimp\)) in the interpretation of separating implication is thus analogous to the combination of (\(\bigvee\)) and (\(\otimes\)) in the interpretation of the separating conjunction (\(*\)).

\begin{example}
    The interpretation of the predicate \(\psi\equiv q_1\mapsto|0\>\<0|\sepimp q_1,q_2\mapsto|01\>\<01|\) on domain \(\{q_2\}\) is given by
    \[
    \begin{aligned}
        \sem{\psi}(\{q_2\})&=\bigwedge_{\substack{\tau\in\Dom\\ q_2\notin \tau}}\sem{q_1\mapsto|0\>\<0|}(\tau)\otimesimp\sem{q_1,q_2\mapsto|01\>\<01|}(\tau\cup\{q_2\})\\
        &=\underbrace{(|0\>_{q_1}\<0|\otimesimp|01\>_{q_1,q_2}\<01|)}_{\tau=\{q_1\}}\wedge \bigwedge_{\substack{q_2\notin\tau\\ \tau\neq\{q_1\}}}\pare{\mathbf{0}_{\tau}\otimesimp \bzero_{\tau \cup\{q_2\}}}\\ &=|1\>_{q_2}\<1|\wedge I_{q_2}=|1\>_{q_2}\<1|,
    \end{aligned}
    \]
    where by Lemma~\ref{lemma:tensor-implication}, \((\bzero_\tau\otimesimp\bzero_{\tau \cup\{q_2\}})=I_{q_2}\) for arbitrary \(\tau\). 
    It can be verified that \(\sem{\psi}(\tau)=\bzero\) for any \(\tau\neq \{q_2\}\).
    Therefore, \(\psi\equiv q_1\mapsto|0\>\<0|\sepimp q_1,q_2\mapsto|01\>\<01|\) is logically equivalent to the assertion \(q_2\mapsto|1\>\<1|\).
    The interpretation suggests that, for the quantum heap \(\rho\), if \(\rho\vDash q_2\mapsto|1\>\<1|\), then after appending another heap satisfying \(q_1\mapsto|0\>\<0|\) the whole heap will satisfy \(q_1,q_2\mapsto|01\>\<01|\). 
    This exactly reflects our intuitive understanding of predicates with separating implications.
\end{example}
\begin{example}
    Consider \(\psi\equiv q\mapsto I\sepimp q\hookrightarrow I\). 
    By careful computation, it can be verified that for any domain \(\tau\in\Dom\), \(\sem{\psi}(\tau)=I\), which means \(\psi\) is logically equivalent to \(\top\).

    Intuitively, \(q\mapsto I\sepimp q\hookrightarrow I\) requires that if a heap is appended with domain \(\{q\}\), then the whole heap will contain at least \(q\). This is obviously a tautology.
\end{example}

\subsection{Reasoning in Quantum Bunched Logic}

In this section, we discuss how to define and reason about entailment between 
predicates in the quantum bunched logic introduced in the previous section. 
Formally, a predicate \(\psi\) is said to \emph{entail} another predicate \(\varphi\), written \(\psi \models \varphi\), if and only if the following condition holds:
\[
\forall \tau\in \Dom,\quad
\sem{\psi}(\tau) \subseteq \sem{\varphi}(\tau).
\]
Here, the subset relation \(\subseteq\) between projection operators coincides with entailment in Birkhoff--von Neumann quantum logic and induces the order on the orthomodular lattice of projections, degenerating to \(\leq\) when \(\tau = \emptyset\). 
This entailment is natural and consistent with the intuitions illustrated below.

\begin{proposition}\label{prop:entailment}
    For arbitrary predicates \(\psi\) and \(\varphi\), it holds that \(\psi \models \varphi\)
    if and only if
    \[
    \forall \rho \in \QHeap,\quad
    \rho \models \psi \implies \rho \models \varphi.
    \]
\end{proposition}
In other words, the entailment \(\psi \models \varphi\) precisely captures the fact that any quantum heap satisfying \(\psi\) must also satisfy \(\varphi\).
\begin{example}\label{eg:entail}
    It can be verified that \(\bar{q} \mapsto P \models \bar{q} \hookrightarrow P\) for arbitrary projection operator \(P\), since for any domain \(\tau\in\Dom\),
    \[
    \sem{\bar{q} \mapsto P}(\tau) =
    \begin{cases}
    P, & \text{if } \bar{q} = \tau;\\
    \bzero, & \text{otherwise},
    \end{cases}
    \quad\mbox{and}\quad 
    \sem{\bar{q} \hookrightarrow P}(\tau) =
    \begin{cases}
    P\otimes I_{\tau \setminus \bar{q}}, & \text{if } \bar{q} \subseteq \tau;\\
    \bzero, & \text{otherwise},
    \end{cases}
    \]
    so that \(\sem{\bar{q} \mapsto P}(\tau) \subseteq \sem{\bar{q} \hookrightarrow P}(\tau)\) for any domain \(\tau\).  
    This entailment aligns with our intuition, as the predicate \(\bar{q} \mapsto P\)
    imposes a stricter condition on the quantum heap---the heap contains exactly \(\bar{q}\)---than \(\bar{q} \hookrightarrow P\), which requires the heap to contain at least \(\bar{q}\).
\end{example}
\begin{figure*}[t]
    \centering
    $
    \begin{array}{lc@{\hspace{3em}}lc@{\hspace{3em}}lc}
        0.&\inferrule[]{\;}{\neg\neg\psi\vdash\psi}
        &1.& \inferrule[]{\;}{\psi\vdash\psi} 
        &2.& \inferrule[]{\;}{\psi\vdash\top} \\
        3.&\inferrule[]{\;}{\bot\vdash\psi}
        &4.&\inferrule[]{\eta\vdash\psi\\ \eta\vdash\varphi}{\eta\vdash\psi\wedge\varphi}
        &5.&\inferrule[]{\eta\vdash\psi_1\wedge\psi_2}{\eta\vdash\psi_i}\\
        6.&\inferrule[]{\varphi\vdash\psi}{\eta\wedge\varphi\vdash\psi}
        &7.&\inferrule[]{\eta\vdash\psi\\ \varphi\vdash\psi}{\eta\vee\varphi\vdash\psi}
        &8.&\inferrule[]{\eta\vdash\psi_i}{\eta\vdash\psi_1\vee\psi_2}\\
        9.&\inferrule[]{\xi\vdash\varphi\\ \eta\vdash\psi}{\xi*\eta\vdash\varphi*\psi}
        &10.&\inferrule[]{\eta\vdash\varphi\sasaki\psi\\ \eta\vdash\varphi}{\eta\vdash\psi}
        &11.&\inferrule[]{\eta\Cap\varphi\vdash\psi}{\eta\vdash\varphi\sasaki\psi}\\
        12.&\inferrule[]{\eta*\varphi\vdash\psi}{\eta\vdash\varphi\sepimp\psi}
        &13.&\inferrule[]{\xi\vdash\varphi\sepimp\psi \\ \eta\vdash\varphi}{\xi*\eta\vdash\psi}
        &14.&\inferrule[]{\;}{(\varphi*\psi)*\xi\vdash\varphi*(\psi*\xi)}\\
        15.&\inferrule[]{\;}{\psi*\varphi\vdash\varphi*\psi}
        &16.&\inferrule[]{\;}{\psi*\top^*\vdash\psi}
        &17.&\inferrule[]{\;}{\psi\vdash\psi*\top^*}
    \end{array}
    $
    \caption{Inference system for quantum bunched logic, where \(i=1,2\) in Rules~5 and~8. The system preserves all inference rules of classical bunched logic as in~\cite{bias, bithesis}.}
    \label{fig:qbi-proofsystem}
\end{figure*}
Next, we present a Hilbert-style proof system for reasoning about entailment
shown in Figure~\ref{fig:qbi-proofsystem}, where \(\vdash\) denotes derivability in the proof system.
Although G\"odel's incompleteness theorem implies that such an inference system can hardly be complete\footnote{We refrain from making a strict claim here, as the arithmetic expressiveness of the logic required by G\"odel's incompleteness theorems has not been formally established; and completeness is typically not a primary concern in the development of program logics.}---namely, that there exist cases where \(\psi \models \varphi\) does not entail \(\psi \vdash \varphi\)---we emphasize that all inference rules from classical bunched logic in~\cite{bias, bithesis} remain valid in our quantum bunched logic, so that existing reasoning tools can be directly applied.

Of particular note, Rules~12 and~13 precisely capture the adjunction between the separating conjunction (\(*\)) and the separating implication (\(\sepimp\)), and provide the formal justification for the first intuition about separating implication discussed in Section~\ref{sec:bilogic-semantics}: namely, that \(\psi_1 \sepimp \psi_2\) is the weakest assertion \(\varphi\) such that \(\psi_1 * \varphi\) entails \(\psi_2\).

We conclude this short section by claiming the soundness of the inference system and presenting an example of reasoning.
\begin{theorem}[Soundness of the inference system]\label{thm:qbi-soundness}
    The inference system presented in Figure~\ref{fig:qbi-proofsystem} is sound: if \(\psi\vdash\varphi\) is derivable from the inference rules, then \(\psi \models \varphi\).
\end{theorem}

\begin{example}
    Continuing from Example~\ref{eg:entail}, we illustrate how to derive \(\bar{q} \mapsto P \vdash \bar{q} \hookrightarrow P\) within the inference system.
    By Rules~1 and~2, we can derive \(\bar{q} \mapsto P \vdash \bar{q} \mapsto P\) and \(\top^* \vdash \top\), respectively.
    Applying Rule~9 then yields
    \[
    \bar{q} \mapsto P * \top^* \vdash \bar{q} \mapsto P * \top.
    \] 
    Finally, since Rule~17 implies \(\bar{q} \mapsto P \vdash \bar{q} \mapsto P * \top^*\), we conclude \(\bar{q} \mapsto P \vdash \bar{q} \hookrightarrow P\).
\end{example}

\section{A Quantum Programming Language with Heap Manipulations}\label{sec:prog}
In this section, we present a formal quantum programming language \textsf{QWhile-Heap} that introduces heap manipulations discussed in Section~\ref{sec:qheap} into the widely-acknowledged language \textsf{QWhile}~\cite{qhoare}, which will serve as the target language for our quantum separation logic in the subsequent section.

\begin{figure}[t]
    \centering
    \[
    \begin{aligned}
        \mbox{(Statement)}\quad  S &::= && \tskip \mid U[\bar{q}]\mid S_1;S_2\mid \tif\ \cM[\bar{q}]\ \tthen\ S_1\ \telse\ S_2\\
        & \ \mid && \twhile\ \cM[\bar{q}]\ \tdo\ S\ \tend\mid q := \talloc(d) \mid \trelease(q)
    \end{aligned}
    \]
    \caption{Syntax of the \textsf{QWhile-Heap} programming language.}
    \label{fig:qwhile-heap-syntax}
\end{figure}

Specifically, the syntax of \textsf{QWhile-Heap} is given in Figure~\ref{fig:qwhile-heap-syntax}, where \(U\) denotes a unitary transformation, \(\cM = \{P_\TT, P_\FF\}\) represents a binary projective quantum measurement with \(P_\TT\) and \(P_\FF\) being projection operators satisfying \(P_\TT + P_\FF = I\), and \(d \ge 2\) is a natural number.
We restrict our attention to projection-valued measurements (PVMs) rather than general positive operator-valued measurements (POVMs) to simplify reasoning in the subsequent separation logic, since the quantum bunched logic introduced earlier is interpreted in terms of projection operators. 
This restriction incurs no loss of generality, as Naimark's dilation theorem~\cite{qcqi} guarantees that any POVM can be realized by allocating ancillary qubits, applying an appropriate unitary transformation, and performing a PVM.

The initialization statement \([q] := |0\>\) in \textsf{QWhile} can be defined as syntactic sugar in \textsf{QWhile-Heap} as
\(
\tif\ \cM[q]\ \tthen\ \tskip\ \telse\ X[q],
\)
where \(q\) points to a qubit, \(\cM=\{P_\TT=|0\rangle\langle 0|,\,P_\FF=|1\rangle\langle 1|\}\), and \(X\) denotes the Pauli-\(X\) operator.
The general initialization of qudits can be defined similarly by performing an appropriate measurement and applying a unitary conditioned on the measurement outcome, such that the resulting state is transformed to \(|0\>\) in all cases.

\begin{figure}[t]
    \centering
    {\small
    \[
    \begin{array}{c}
        \inferrule[]{\;}{(\tskip,\rho)\to (\downarrow,\rho)}\quad 
        \inferrule[]{\bar{q}\subseteq\dom\rho}{(U[\bar{q}],\rho)\to (\downarrow,U_{\bar{q}}\rho U_{\bar{q}}^\dagger)}\quad
        \inferrule[]{(S_1,\rho)\to (\downarrow,\rho')}{(S_1;S_2,\rho)\to (S_2,\rho')}\quad 
        \inferrule[]{(S_1,\rho)\to (S_1',\rho')}{(S_1;S_2,\rho)\to (S_1';S_2,\rho')} \\~\\
        \inferrule[]{\bar{q}\subseteq \dom\rho \\ \cM=\{P_\TT,P_\FF\}}{(\tif\ \cM[\bar{q}]\ \tthen\ S_1\ \telse\ S_2,\rho)\to (S_1,P_\TT\rho P_\TT)}\qquad 
        \inferrule[]{\bar{q}\subseteq \dom\rho \\ \cM=\{P_\TT,P_\FF\}}{(\tif\ \cM[\bar{q}]\ \tthen\ S_1\ \telse\ S_2,\rho)\to (S_2,P_\FF\rho P_\FF)}\\~\\
        \inferrule[]{\bar{q}\subseteq \dom\rho \\ \cM=\{P_\TT,P_\FF\}}{(\twhile\ \cM[\bar{q}]\ \tdo\ S\ \tend,\rho)\to (S;\twhile\ \cM[\bar{q}]\ \tdo\ S\ \tend,P_\TT\rho P_\TT)}\quad
        \inferrule[]{\bar{q}\subseteq \dom\rho \\ \cM=\{P_\TT,P_\FF\}}{(\twhile\ \cM[\bar{q}]\ \tdo\ S\ \tend,\rho)\to (\downarrow,P_\FF\rho P_\FF)} \\~\\
        \inferrule[]{\rho'\in \cD(\cH_{\dom\rho[\sqcup/q]}\otimes \cH_{q})\quad \dim\cH_q=d\quad \Tr_q(\rho')=\rho[\sqcup/q]}{(q:=\talloc(d),\rho)\to (\downarrow,\rho')}\quad
        \inferrule[]{q\in \dom\rho}{(\trelease(q),\rho)\to (\downarrow,\Tr_q(\rho))}
    \end{array}
    \]}
    \caption{Small-step operational semantics of \textsf{QWhile-Heap}.}
    \label{fig:qwhile-heap-semantics}
\end{figure}

A \emph{program configuration} in \textsf{QWhile-Heap} is a pair \((S, \rho) \in \mathsf{Config} \triangleq (\mathsf{Statement} \cup \{\downarrow\}) \times \QHeap\),  where the first component denotes either a program statement \(S\) to be executed or the special symbol ``\(\downarrow\)'' indicating termination, and the second component represents the current quantum heap. 
The small-step operational semantics of \textsf{QWhile-Heap} is defined as a transition relation 
\(\to \;\subseteq\; \mathsf{Config} \times \mathsf{Config}\), 
where \((S_1, \rho_1) \to (S_2, \rho_2)\) denotes a single execution step (as an alternative to the set-theoretic notation \(((S_1, \rho_1), (S_2, \rho_2)) \in \to\)). 
The relation \(\to\) is determined by the inference rules given in Figure~\ref{fig:qwhile-heap-semantics}. 
Furthermore, we write \(\to^* = \bigcup_{n=0}^\infty \to^n\) for the reflexive and transitive closure of \(\to\).

Quantum programs inherently exhibit probabilistic behavior because branching and looping constructs are controlled by measurement outcomes.
For an execution path \((S,\rho)\!\to\!(S_1,\rho_1)\!\to\cdots\to\!(S',\rho')\) with \(\Tr(\rho)=1\), the trace of the final state \(\Tr(\rho')\) represents the probability of taking all branches along that path.
In addition, as discussed in Section~\ref{sec:qheap}, the allocation statement \(q:=\talloc(d)\) introduces nondeterminism, since it can result in various possible quantum heaps depending on the state of the newly allocated qudit.

Next, we will provide more detailed explanations and examples of some rules.

\paragraph{Accessing Heaps: \(U[\bar{q}],\ \cM[\bar{q}]\)} 
The unitary statement \(U[\bar{q}]\) applies the unitary operator \(U\) to the qudits referenced by the variables in \(\bar{q}\), as described in Section~\ref{sec:qheap}.
Similarly, the measurements guarding branching and looping statements act on the
qudits referenced by \(\bar{q}\), potentially modifying the quantum heap---unlike
classical branching where the guard condition is only evaluated without side effects.

Brackets are used to emphasize that operations are performed on the qudits pointed to by the variables, rather than on the variables themselves.
If \(\bar{q} \subseteq \dom \rho\) does not hold, no rule applies and the execution is said to be \emph{stuck}, following the classical terminology \cite{bias}.
In the next section, we show that an important property of the separation logic is that the satisfaction of the preconditions in program specifications guarantees that programs do not get stuck.

\paragraph{Allocation and Deallocation: \(q:=\talloc(d),\ \trelease(q)\)}
The allocation statement \(q:=\talloc(d)\) can be understood as first allocating a new \(d\)-dimensional qudit in an arbitrary state and then letting the variable \(q\) point to it, as discussed in Section~\ref{sec:qheap}.
The dimension of the qudit reflects its type, e.g., \(d=2\) for qubits of Boolean type~\cite{Ying24}, and we do not consider infinite-dimensional types to avoid technical complications in this work.

In the allocation rule, we write \(\rho[\sqcup/q]\) for the quantum heap obtained by replacing every occurrence of \(q\) in the subscripts of \(\rho\) with ``\(\sqcup\)''.
The only requirement on the resulting heap \(\rho'\) is that tracing out the newly allocated qudit yields the original heap, allowing flexible nondeterministic choices for the state of the new qudit while preserving probability, i.e., \(\Tr(\rho') = \Tr(\rho)\).

The deallocation rule for \(\trelease(q)\) is straightforward: it simply traces out the qudit referenced by \(q\) from the quantum heap. 
However, if the target variable has not been allocated, i.e., \(q \notin \dom \rho\), the deallocation becomes stuck.

\section{A Quantum Separation Logic}\label{sec:separation}
Separation logic is a variant of Hoare logic that adopts bunched logic as its assertion logic to support reasoning about heap-manipulating programs~\cite{bias,reynolds}.
In this section, we develop a quantum separation logic for reasoning about the quantum programs introduced in Section~\ref{sec:prog}, using the quantum bunched logic from Section~\ref{sec:bilogic} as the assertion logic.

Formally, properties of \textsf{QWhile-Heap} quantum programs are specified by Hoare-style triples \(\triple{\psi}{S}{\varphi}\), where \(S\) is a program statement of the \textsf{QWhile-Heap} language (Figure~\ref{fig:qwhile-heap-syntax}), and \(\psi\) and \(\varphi\) are predicate formulas in the quantum bunched logic (Figure~\ref{fig:qbi-formula}).
For partial correctness---where termination is not required---the triple holds, denoted \(\models \triple{\psi}{S}{\varphi}\), if and only if the following two
conditions are satisfied:
\begin{enumerate}
    \item for any quantum heap \(\rho \in \QHeap\) satisfying the
    precondition \(\psi\) (i.e., \(\rho \models \psi\)), the execution of \(S\)
    on \(\rho\) never gets stuck: for any transition
    \((S,\rho) \to^* (S',\rho')\), either \(S'=\,\downarrow\) or there exists a
    further transition \((S',\rho') \to (S'',\rho'')\); and
    \item for any quantum heap \(\rho \in \QHeap\) satisfying the
    precondition \(\psi\), the execution of \(S\) on \(\rho\) terminates only in
    heaps satisfying the postcondition \(\varphi\): for any
    \((S,\rho) \to^* (\downarrow,\rho')\), we have \(\rho' \models \varphi\).
\end{enumerate}

It is worth noting that, unlike Hoare logic---where only the second condition is required for partial correctness---separation logic imposes stronger requirements on the precondition to ensure that all accessed memory locations are properly allocated, preventing the program from getting stuck during execution. 
Accordingly, reasoning about program specifications will also involve reasoning about whether a program may get stuck.
This practice is standard in classical separation logic but has not been adopted in previous quantum separation logics~\cite{qsep,qsep1,qsep2,rapunsl}.

We first present the proof system consisting of the \emph{small axioms}~\cite{separationlogic}, and illustrate that the proof system is sound, relatively complete, and consistent with our intuitive understanding.
Next, we will introduce backward-reasoning rules with general postconditions and separating implications in preconditions.

\subsection{Proof System with Small Axioms}

\begin{figure*}
    \centering
    $
    \begin{array}{c}
        \multicolumn{1}{l}{\large\boxed{Small\ Axioms}} \\

        \inferrule{\;}{\triple{\psi}{\tskip}{\psi}}{\textsc{Skip}}
        
        \hspace{5.0em}

        \inferrule{\bar{q}\subseteq \bar{q}'\subseteq_{\mathrm{fin}}\mathsf{Var}\\ P\in\cP(\cH_{\bar{q}'})}{\triple{\bar{q}'\mapsto P}{U[\bar{q}]}{\bar{q}'\mapsto U_{\bar{q}}PU_{\bar{q}}^\dagger}}{\textsc{Unitary}} 
        
        \\[1em]
        
        \inferrule{\;}{\triple{\top^*}{q:=\talloc(d)}{q\mapsto I}}{\textsc{Allocate}}
        
        \hspace{3.0em}

        \inferrule{\;}{\triple{q\mapsto I}{\trelease(q)}{\top^*}}{\textsc{Release}}
        
        \\[2em]

        \multicolumn{1}{l}{\large\boxed{Compound\ Statements}} \\[1em]
        
        \inferrule{\triple{\psi}{S_1}{\varphi}\\ \triple{\varphi}{S_2}{\phi}}{\triple{\psi}{S_1;S_2}{\phi}}{\textsc{Sequence}}
        
        \hspace{3.0em}
        
        \inferrule{\psi \models\psi'\\ \triple{\psi'}{S}{\varphi'}\\ \varphi' \models\varphi}{\triple{\psi}{S}{\varphi}}{\textsc{Consequence}}
        
        \\[1.5em] 
        
        \inferrule{\cM=\{P_\TT,P_\FF\}\\ \triple{\psi_\TT}{S_1}{\psi}\\ \triple{\psi_\FF}{S_2}{\psi}}{\triple{(\bar{q}\hookrightarrow I)\wedge\bigwedge\nolimits_{m\in\{\TT,\FF\}}(\bar{q}\hookrightarrow P_m)\sasaki \psi_m}{\tif\ \cM[\bar{q}]\ \tthen\ S_1\ \telse\ S_2}{\psi}}{\textsc{If}}

        \\[2.5em]
        
        \inferrule
        {\cM=\{P_\TT,P_\FF\}\\
         \triple{\varphi}{S}{\phi}\\ \phi\equiv (\bar{q}\hookrightarrow I)\wedge((\bar{q}\hookrightarrow P_\FF)\sasaki \psi)\wedge ((\bar{q}\hookrightarrow P_\TT)\sasaki \varphi)}
         {\triple{\phi}{\twhile\ \cM[\bar{q}]\ \tdo\ S\ \tend}{\psi}}{\textsc{While}}\\[2em]

        \multicolumn{1}{l}{\large\boxed{Structural\ Rules}} \\[1em]

        \inferrule[\textsc{Conjunction}]{\triple{\psi_1}{S}{\varphi_1}\ \ \ \triple{\psi_2}{S}{\varphi_2}}{\triple{\psi_1\wedge\psi_2}{S}{\varphi_1\wedge\varphi_2}}

        \quad    
        
        \inferrule[\textsc{Disjunction}]{\triple{\psi_1}{S}{\varphi_1}\ \ \ \triple{\psi_2}{S}{\varphi_2}}{\triple{\psi_1\vee\psi_2}{S}{\varphi_1\vee\varphi_2}}
        
        \quad    

        \inferrule[\textsc{Frame}]{\triple{\psi}{S}{\varphi}\ \ \ \mathrm{var}(\phi)\cap \mathrm{mod}(S)=\emptyset}{\triple{\psi*\phi}{S}{\varphi*\phi}}
    \end{array}
    $
    \caption{Proof system for quantum separation logic with small axioms.}
    \label{fig:qsep-proof-system}
\end{figure*}

Figure~\ref{fig:qsep-proof-system} presents the proof system of our quantum separation logic, where the inference rules are grouped into three categories:
\begin{enumerate}
    \item \emph{Small Axioms}~\cite{separationlogic} for primitive program statements, which specify how locally modified heaps are handled without referring to the global heap.
    \item \emph{Compound Statements} for reasoning about sequential composition, branching, and looping constructs, based on a structural induction on program statements.
    \item \emph{Structural Rules} for manipulating preconditions and postconditions without altering the program statements.
\end{enumerate}

We next examine each rule—except for the trivial \textsc{Skip}—in detail and discuss its intended meaning.

\paragraph{Small axioms}
Analogous to the assertion \(x\mapsto -\) in classical separation logic, the assertion \(q\mapsto I\) merely indicates the existence of the qudit referenced by variable \(q\) without specifying its state, since the identity operator \(I\) is interpreted as the projector onto the entire space.  
Moreover, \(\top^*\) asserts an empty heap, similar to the assertion \(\mathtt{emp}\) in classical separation logics~\cite{bias}.  

The inference rules \textsc{Allocate} and \textsc{Release} can be considered as a paired set: \textsc{Allocate} states that allocating a new variable \(q\) transforms an empty heap into a heap containing only \(q\) in an arbitrary state, while \textsc{Release} states that releasing \(q\) transforms a heap containing only \(q\) in an arbitrary state back into an empty heap.

Such small axioms describe only a local portion of the heap---precisely the single variable directly affected by the allocation or release operation---without compromising completeness, as they can be combined with the structural rules introduced later to handle larger heaps.

However, the rule \textsc{Unitary} for unitary transformations differs from classical assignment rules.  
Unlike the classical assignment rule \(\triple{x\mapsto -}{[x]:=v}{x\mapsto v}\), which accounts for a heap containing exactly one memory cell with address \(x\), the rule for unitary transformations cannot be restricted to heaps containing only the variables in \(\bar{q}\), i.e.,  \(\triple{\bar{q}\mapsto P}{U[\bar{q}]}{\bar{q}\mapsto UPU^\dagger}\), as doing so would render it incomplete.  

This incompleteness arises because, in the quantum setting, a global heap contains more information than the sum of its individual components due to inter-system correlations such as entanglement, 
whereas a classical heap can be reconstructed perfectly from its constituent pieces.  
Consequently, larger heaps must be taken into account, and the rule \textsc{Unitary} is formulated to handle heaps containing at least \(\bar{q}\) to ensure completeness.
In contrast, the small axioms for allocation and deallocation do not encounter this issue, since partial trace and tensoring with the identity are inherently entanglement-free.

\paragraph{Rules for compound statements} Rules \textsc{Sequence} and \textsc{Consequence} are standard in Hoare logic and separation logic~\cite{bias}. 
In particular, the entailments \(\psi \models \psi'\) and \(\varphi' \models \varphi\) used in \textsc{Consequence} are defined in Section~\ref{sec:bilogic}, 
and the reasoning about these entailments is discussed in Section~\ref{sec:bilogic} and carried out independently from the reasoning about program specifications.

Rules \textsc{If} and \textsc{While} for branching and looping are more subtle than their classical counterparts, as their guards are quantum measurements, which themselves are quantum operations.
To ensure that a measurement never gets stuck due to missing target qudits, the preconditions in both rules include a conjunct of the form \(\bar{q} \hookrightarrow I\), guaranteeing the existence of the qudits \(\bar{q}\) being measured.

The Sasaki hook \((\bar{q} \hookrightarrow P_m) \sasaki \psi_m\) serves as a quantum analogue of logical implication in classical Hoare logic, expressing that \emph{if} the measurement outcome is \(m\) (i.e., \(\bar{q} \hookrightarrow P_m\)), \emph{then} the predicate \(\psi_m\) must hold after the measurement.
The predicate \(\phi\) in \textsc{While} is often regarded as the \emph{loop invariant}, since it is preserved before and after each iteration of the loop.

The inference rule for \textbf{if} statements can be presented in another form:
\[
\inferrule{\cM=\{P_\TT,P_\FF\}\\ \triple{\psi_1}{S_1}{\psi}\\ \triple{\psi_2}{S_2}{\psi}}{\triple{\bigvee\nolimits_{m\in\{\TT,\FF\}}(\bar{q}\hookrightarrow P_m)\wedge\psi_m}{\tif\ \cM[\bar{q}]\ \tthen\ S_1\ \telse\ S_2}{\psi}}{\textsc{If}}.
\]
Such an inference rule appears to be tidier and more concise as the precondition inherently implies the existence of \(\bar{q}\).
However, it is discouraged in practice because the (\(\vee\))-introduction rule (i.e., Rule~8 in Figure~\ref{fig:qbi-proofsystem}) is very weak in quantum logic.
\(\psi \models \bigvee\nolimits_{m\in\{\TT,\FF\}}(\bar{q}\hookrightarrow P_m)\wedge\psi_m\) can be proved with the inference rules in Figure~\ref{fig:qbi-proofsystem} only if \(\psi \models(\bar{q}\hookrightarrow P_m)\wedge\psi_m\) for some \(m\in\{\TT,\FF\}\), which means that only one branch of the \textbf{if} statement can potentially be executed.

Such a scenario is reasonable in classical programming languages, where guard conditions are mutually exclusive. 
Nevertheless, in quantum programming, multiple branches of the \textbf{if} statements may be taken with nonzero probabilities.
Therefore, the precondition needs to maintain a more easily deducible form, and the Sasaki hook is a well-studied connective with many algebraic properties that aid in reasoning about related formulas.

\paragraph{Structural rules} The rules \textsc{Conjunction} and \textsc{Disjunction} behave exactly as in classical program logics.
It is worth noting that the rule for conjunction does not incur any loss of completeness, since conjunction in quantum logic coincides with the classical one; that is,
\[
\rho \models \psi_1 \wedge \psi_2 
\iff \rho \models \psi_1 \ \text{and}\ \rho \models \psi_2.
\]
Consequently, the weakest precondition with respect to a conjunction is simply the conjunction of the weakest preconditions.

However, unlike in classical program logics, the disjunction rule compromises completeness in quantum logic, because satisfying a disjunction does not necessarily imply satisfying either of its disjuncts. 
Consequently, the weakest precondition with respect to a disjunction may be substantially weaker than the disjunction of the weakest preconditions, as illustrated in the following example.
\begin{example}[Incompleteness of \textsc{Disjunction}]
    Recall that for a variable \(q\) pointing to a qubit, we introduced the syntactic sugar
    \[
    [q] := |0\rangle \;\equiv\; 
    \tif\ \cM_{01}[q]\ \tthen\ \tskip\ \telse\ X[q]
    \]
    to initialize \(q\) to the state \(|0\>\), where 
    \(\cM_{01}=\{P_\TT=|0\>\< 0|,P_\FF=|1\>\< 1|\}\) is the computational-basis measurement.
    It is then straightforward to verify that the following two triples are valid:
    \[
    \triple{\bot}{[q]:=|0\>}{q \mapsto |+\>\< +|},
    \ \ 
    \triple{\bot}{[q]:=|0\>}{q \mapsto |-\>\<-|}.
    \]
    The precondition \(\bot\) is the weakest possible for both triples, since once \(q\) is initialized to \(|0\>\), it is impossible to reach either \(|+\>\) or \(|-\>\).

    Consequently, applying the \textsc{Disjunction} rule yields
    \[
    \triple{\bot}{[q]:=|0\>}
        {(q \mapsto |+\>\< +|)\vee(q \mapsto |-\>\< -|)}.
    \]
    However, the precondition \(\bot\) is far from the weakest in this case, because the postcondition is equivalent to \(q \mapsto I\) in quantum logic, and thus \(q \mapsto I\) is also a valid precondition—strictly weaker than \(\bot\) as \(\bot\models q\mapsto I\).
\end{example}

Rule \textsc{Frame} takes the same form as in classical separation logic, where \(\mathrm{var}(\phi)\) denotes the variables occurring free in the predicate \(\phi\) and \(\mathrm{mod}(S)\) denotes the set of stack variables that may be modified by the statement \(S\).  
Since quantifiers are not included in our quantum bunched logic, all variables are considered free in any predicate.  
Moreover, the \textsf{QWhile-Heap} language does not support variable assignments of the form \(q:=e\); hence, the only way a variable can be modified is through allocation, \(q:=\talloc(d)\). 

We illustrate, by means of a concrete example, why the condition \(\mathrm{mod}(S)\cap \mathrm{var}(\phi)=\emptyset\) is sufficient to guarantee the soundness of \textsc{Frame}, without imposing any additional requirement that the target variables of \(S\) (e.g., \(\bar{q}\) in \(S\equiv U[\bar{q}]\)) be disjoint from \(\mathrm{var}(\phi)\), as one might otherwise expect.
Consider the statement \(S\equiv U[q]\) with \(\mathrm{mod}(S)=\emptyset\).
In this case, a valid specification of \(S\) is \(\triple{q\mapsto I}{U[q]}{q\mapsto I}\).
Now suppose another predicate \(\phi\equiv q\mapsto P\) involving \(q\) is introduced.
Then the precondition \(q\mapsto I * q\mapsto P\) is satisfiable only by zero heaps, and hence
the specification \(\triple{q\mapsto I * q\mapsto P}{U[q]}{q\mapsto I * q\mapsto P}\) becomes trivially valid.
Therefore, the first requirement of the program specification, which ensures that satisfaction of the precondition implies that the program does not get stuck, plays a crucial role in guaranteeing the soundness of \textsc{Frame}, as mentioned in Section~\ref{sec:introduction}.

Besides, rule \textsc{Frame} is central to local reasoning, as it specifies the purpose of separating conjunction \((*)\).
Unlike structural rules for conjunction and disjunction, the premise of \textsc{Frame} concerns only a single component of the separating conjunction \((*)\).
This permits reasoning to be performed on an individual component of the predicate, as shown in the following example.

\begin{example}\label{eg:frame}
    To derive the valid triple
    \[
    \triple{q_1\mapsto|0\>\<0|*q_2\mapsto|0\>\<0|}{H[q_1]}{q_1\mapsto|+\>\<+|*q_2\mapsto|0\>\<0|},
    \]
    variable \(q_2\) in the postcondition can be temporarily ignored and \(\triple{q_1\mapsto|0\>\<0|}{H[q_1]}{q_1\mapsto|+\>\<+|}\) is proved with the small axiom \textsc{Unitary}.  
    The full result is then obtained using rule \textsc{Frame} by simply adding \(q_2\mapsto|0\>\<0|\) without any additional burden.
\end{example}

We conclude this subsection by stating the soundness and relative completeness of the proof system.
\begin{theorem}[Soundness and Relative Completeness]\label{thm:sep-sound-complete}
    The proof system presented in Figure~\ref{fig:qsep-proof-system} is 
    \begin{enumerate}
        \item \emph{Sound}: if \(\triple{\psi}{S}{\varphi}\) is derivable from the inference rules, then \(\models \triple{\psi}{S}{\varphi}\) holds.
        \item \emph{Relatively Complete}: if \(\models \triple{\psi}{S}{\varphi}\) holds, then \(\triple{\psi}{S}{\varphi}\) is derivable from the inference rules, assuming that all entailments \(\psi\models\varphi\) between predicates can be decided.
    \end{enumerate}
\end{theorem}
The relative completeness result obtained here lies between the extensional and intensional viewpoints.
We do not analyze the syntax or formula structure of the underlying logic; instead, we simply assume that all projection operators are available as primitive predicates of the form \(\bar{q} \mapsto P\).
The key step in establishing relative completeness is to reduce the domain-sensitive interpretation of predicates with bunched logical connectives to the completeness of the underlying quantum logic generated by all projection operators. 
This reduction is feasible because each predicate formula involves only finitely many variables, and its interpretation over arbitrary domains is determined by finitely many canonical domain configurations, together with their cylinder extensions obtained by tensoring with identity operators on irrelevant variables.

\subsection{Backward-reasoning Rules with General Postconditions}
Next, we will present alternative rules for small axioms in Figure~\ref{fig:qsep-proof-system}, which support backward reasoning with general postconditions and involve separating implication (\(\sepimp\)) in preconditions.
\paragraph{Unitary Transformation}
The inference rule is given by
\begin{equation}\label{eq:backward-unitary}
    \inferrule{
        \bar{q}\subseteq\bar{q}'\subseteq_{\mathrm{fin}}\mathsf{Var} \\
        P\in\cP(\cH_{\bar{q}'})
    }{
        \triple{
            \bar{q}'\mapsto P * 
            (\bar{q}'\mapsto U_{\bar{q}} P U_{\bar{q}}^\dagger \sepimp \psi)
        }{
            U[\bar{q}]
        }{
            \psi
        }
    }{\textsc{Unitary}},
\end{equation}
where the conclusion triple contains a general postcondition~\(\psi\).

The precondition  \(\bar{q}'\mapsto P * (\bar{q}'\mapsto U_{\bar{q}} P U_{\bar{q}}^\dagger \sepimp \psi )\) admits an intuitive interpretation.  
For any state \(\rho\) satisfying this precondition, \(\rho\) contains a component over \(\bar{q}'\) which satisfies \(P\).
The remaining part of \(\rho\), outside \(\bar{q}'\), satisfies the predicate \(\bar{q}'\mapsto U_{\bar{q}}PU_{\bar{q}}^\dagger\sepimp\psi\), which means that if this remaining part is extended with \(\bar{q}'\) satisfying \(U_{\bar{q}} P U_{\bar{q}}^\dagger\), then the resulting state will satisfy \(\psi\).
Therefore, in a ``remove-then-reintroduce'' manner, the precondition ensures that the unitary transformation does not get stuck (since any state satisfying the precondition must contain \(\bar{q}'\) and \(\bar{q}\)), and that after applying \(U\) to \(\bar{q}\), the resulting state satisfies \(\psi\).

Such a backward-reasoning rule is equivalent to the corresponding small axiom.  
For one direction, by instantiating the predicate \(\psi\) in Eq.~\eqref{eq:backward-unitary} as  \(\bar{q}' \mapsto U_{\bar{q}} P U_{\bar{q}}^\dagger\), the backward-reasoning rule yields
\[
\triple{\bar{q}'\mapsto P * (\bar{q}'\mapsto U_{\bar{q}}PU_{\bar{q}}^\dagger \sepimp \bar{q}'\mapsto U_{\bar{q}}PU_{\bar{q}}^\dagger)}{U[\bar{q}]}{\bar{q}'\mapsto U_{\bar{q}}PU_{\bar{q}}^\dagger}.
\]
Thus, by rule \textsc{Consequence}, it suffices to show that
\[
\bar{q}'\mapsto P\models\bar{q}'\mapsto P *(\bar{q}'\mapsto U_{\bar{q}}PU_{\bar{q}}^\dagger\sepimp \bar{q}'\mapsto U_{\bar{q}}PU_{\bar{q}}^\dagger).
\]
This entailment follows directly from the sound proof system in Figure~\ref{fig:qbi-proofsystem}, simply by observing that
\[
\top^* \models \bar{q}'\mapsto U_{\bar{q}}PU_{\bar{q}}^\dagger\sepimp \bar{q}'\mapsto U_{\bar{q}}PU_{\bar{q}}^\dagger.
\]
Therefore, from the backward-reasoning rule, we can derive the small axiom for unitary transformations.

For the converse direction, one may apply rule \textsc{Frame} to attach the predicate \((\bar{q}'\mapsto U_{\bar{q}} P U_{\bar{q}}^\dagger \sepimp \psi)\) to the small axiom, obtaining
\[
\triple{\bar{q}'\mapsto P * (\bar{q}'\mapsto U_{\bar{q}}PU_{\bar{q}}^\dagger \sepimp \psi)}{U[\bar{q}]}{\bar{q}'\mapsto U_{\bar{q}}PU_{\bar{q}}^\dagger * (\bar{q}'\mapsto U_{\bar{q}}PU_{\bar{q}}^\dagger \sepimp \psi)}.
\]
Using the sound proof system in Figure~\ref{fig:qbi-proofsystem}, one can derive
\[
\bar{q}'\mapsto U_{\bar{q}}PU_{\bar{q}}^\dagger *(\bar{q}'\mapsto U_{\bar{q}}PU_{\bar{q}}^\dagger \sepimp \psi)\models\psi,
\]
which, together with rule \textsc{Consequence}, establishes the backward-reasoning rule.

\paragraph{Deallocation} The backward-reasoning rule for deallocation is quite straightforward:
\vspace{-1em}
\begin{equation}\label{eq:backward-release}
\inferrule{\;}{\triple{q\mapsto I*\psi}{\trelease(q)}{\psi}}{\textsc{Release}}.
\end{equation}
The precondition states that the heap contains \(q\) so that the deallocation does not get stuck, and the remaining part of the heap satisfies \(\psi\), which is precisely the postcondition.
It can be verified that the backward-reasoning rule is equivalent to the small axiom for deallocation using similar techniques as for unitary transformations.

\paragraph{Allocation} The backward-reasoning rule for allocation is more complicated.
Similar to the rules in classical separation logic~\cite{bias}, another universal quantifier over variables needs to be introduced into the assertion logic to express the precondition for a general postcondition.
Specifically, we introduce the construct \(\forall q\,\psi\) in the quantum bunched logic, which is interpreted as 
\[
\sem{\forall q\,\psi}(\tau)=\bigwedge\nolimits_{q'\in \mathsf{Var}}\sem{\psi[q'/q]}(\tau).
\]
In other words, such a quantifier ranges over all possible variables \(q'\in\mathsf{Var}\) and takes the conjunction of the interpretations after substituting \(q\) with each variable \(q'\).
\begin{example}
    For the predicate \(\forall q\,(q\mapsto I\sepimp q\hookrightarrow I)\), it can be verified that for arbitrary domain \(\tau\in\Dom\),
    \[
    \begin{aligned}
        \sem{\forall q\,(q\mapsto I\sepimp q\hookrightarrow I)}(\tau)&=\bigwedge\nolimits_{q'\in \mathsf{Var}}\sem{q'\mapsto I\sepimp q'\hookrightarrow I}(\tau)
        =I_\tau.
    \end{aligned}
    \]
    Thus, \(\forall q\,(q\mapsto I\sepimp q\hookrightarrow I)\) is logically equivalent to \(\top\).
\end{example}
The backward-reasoning rule for allocation is then given in a form similar to classical separation logic:
\vspace{-1em}
\begin{equation}\label{eq:backward-allocate}
\inferrule{\;}{\triple{\forall q\,(q\mapsto I\sepimp \psi)}{q:=\talloc(d)}{\psi}}{\textsc{Allocate}}
\end{equation}
It is necessary to introduce (\(\forall\)) because the predicate \(q \mapsto I \sepimp \psi\) degenerates to  \(\top\) on any domain containing \(q\), similar to the classical setting. 
Consequently, every quantum heap that includes \(q\) trivially satisfies \(q \mapsto I \sepimp \psi\).
However, it is evident that not all heaps containing \(q\) will satisfy the postcondition \(\psi\) after the allocation of \(q\), which leads to a loss of soundness.

This deficiency can be addressed by strengthening the precondition to account for all possible variables.  
Intuitively, allocating a fresh qubit \(q\) to a quantum heap \(\rho\) can be viewed as a three-step process:
(1) choose a quantum variable \(q'\notin\dom\rho\);
(2) extend \(\rho\) to include \(q'\);
(3) rename \(q\) to \(\sqcup\), and then rename \(q'\) to \(q\).

By the definition of the universal quantifier, the precondition \(\forall q\, (q \mapsto I \sepimp \psi)\) guarantees that, for the choice of \(q'\) in step~(1), the heap satisfies \(q' \mapsto I \sepimp \psi[q'/q]\).
By the definition of separating implication, after extending \(\rho\) with \(q'\) in step~(2), the heap must satisfy \(\psi[q'/q]\) as \(q'\) is not contained in \(\rho\).
In step~(3), renaming \(q\) to \(\sqcup\) does not affect the satisfaction of \(\psi[q'/q]\), because \(q\) does not appear in \(\psi[q'/q]\) and hence its interpretation does not distinguish between \(q\) and \(\sqcup\).
Finally, renaming \(q'\) to \(q\) transfers the satisfaction of \(\psi[q'/q]\) to the postcondition \(\psi\).
Moreover, the strengthened precondition does not compromise completeness, since instantiating \(\psi\) with \(q \mapsto I\) makes the precondition \(\forall q\, (q \mapsto I \sepimp \psi)\) logically equivalent to \(\top^*\), thereby recovering the small axiom for allocation.



Finally, we conclude that the backward-reasoning rules constitute viable alternatives to the small axioms in Figure~\ref{fig:qsep-proof-system}, since the small axioms can be derived from these backward-reasoning rules.

\begin{theorem}\label{thm:backward-reasoning}

    The small axioms shown in Figure~\ref{fig:qsep-proof-system} can be derived from the backward-reasoning rules in Eqs.~\eqref{eq:backward-unitary}, \eqref{eq:backward-release}, and \eqref{eq:backward-allocate}, together with the structural rules in Figure~\ref{fig:qsep-proof-system}.
\end{theorem}

The theorem indicates that the backward-reasoning rules are sufficient to derive all valid program specifications whose assertions are formulated in the quantum bunched logic introduced in Section~\ref{sec:bilogic}.
More specifically, the backward-reasoning rules for unitary transformations (Eq.~\eqref{eq:backward-unitary}) and deallocation (Eq.~\eqref{eq:backward-release}) are derivable from the small axioms.
By contrast, deriving the backward-reasoning rule for allocation (Eq.~\eqref{eq:backward-allocate}) from the small axioms requires additional structural rules for universal quantification.

Achieving relative completeness for the assertion logic extended with the universal quantifier would, in turn, necessitate further enrichment of both the assertion logic and the structural inference rules---for instance, by admitting infinite conjunctions and disjunctions.
To keep the presentation concise, we do not pursue these additional extensions in the present work.

\section{Extension with Classical Variables}\label{sec:classical}
In this section, we discuss how to extend our framework---including the quantum bunched logic, the programming language, and the separation logic---with classical variables. 
This extension enables reasoning about richer control-flow constructs, as well as properties such as disjointness and classical pointer aliasing.

We present only selected nontrivial components to illustrate the main ideas of the extension; the remaining details can be defined in a straightforward manner and are deferred to the appendix.

\paragraph{Quantum Bunched Logic with Classical Variables}
Instead of referring to qubits by fixed variable names such as \(q_1,q_2,q_3,\ldots\) as defined in Section~\ref{sec:bilogic}, we can incorporate classical variables and allow quantum variables to be indexed by classical expressions. 
Formally, we extend the syntax in Figure~\ref{fig:qbi-formula} by allowing 
\(
\bar{q}=q_1[e_1],\ldots,q_n[e_n]
\),
where each \(e_i\) is a classical expression. We additionally admit Boolean expressions \(b\) as primitive predicates:
\[
\psi ::= \top \mid \bot \mid b \mid \bar{q}\mapsto P \mid \cdots.
\]
Let \(\Sigma:=\mathsf{cVar}\rightharpoonup\mathbb{N}\) denote the set of classical states, where each \(\sigma\in\Sigma\) maps a set of classical variables to natural numbers.
For every formula \(\psi\), we write \(\sigma(\psi)\) for the result of evaluating all classical expressions in \(\psi\) under \(\sigma\), yielding a formula in the original syntax of Figure~\ref{fig:qbi-formula} without classical variables.
The evaluation is defined inductively, with the nontrivial cases given by
\[
\sigma(b)=
\begin{cases}
\top, & \text{if } \sem{b}^\sigma=\TT,\\
\bot, & \text{otherwise},
\end{cases}
\quad\mbox{and}\quad
\sigma(\bar{q}\mapsto P)=
\begin{cases}
\sem{\bar{q}}^{\sigma}\mapsto P,
& \text{if } \sem{\bar{q}}^{\sigma} \mbox{ is pairwise distinct and}\\[-0.3em]
& \mbox{matches the dimension of } P,\\
\bot,
& \text{otherwise}.
\end{cases}
\]
Here \(\sem{b}^\sigma\in\{\TT,\FF\}\) denotes the evaluation of the Boolean expression \(b\) under the classical state \(\sigma\), and \(\sem{\bar{q}}^{\sigma}\) denotes the tuple of quantum variables obtained by evaluating all classical expressions occurring in \(\bar{q}\) under \(\sigma\). 
In the following, we leave the pairwise-distinctness and arity requirements implicit whenever \(\sem{\bar{q}}^{\sigma}\) is used.
We then naturally define the satisfaction relation for formulas with classical variables as
\[
\sigma,\rho\models\psi\iff \rho\models\sigma(\psi).
\]
\begin{example}
Consider the formula
\(
\psi \equiv (x>3)\wedge (q[x]\mapsto |0\>\< 0|).
\)
Let \(\sigma\) be a classical state satisfying \(\sigma(x)=5\). Then \(\sem{q[x]}^\sigma=q[5]\) and
\(
\sigma(\psi)
=
\top\wedge \bigl(q[5]\mapsto |0\>\< 0|\bigr)=q[5]\mapsto |0\>\< 0|.
\)
Therefore, we have 
\[
\sigma,|0\>_{q[5]}\<0|\models (x>3)\wedge (q[x]\mapsto |0\>\< 0|).
\]
\end{example}

\paragraph{Programming Language with Classical Variables}
For the programming language, we similarly allow the target qubits \(\bar{q}\) of unitary operations and measurements to be indexed by classical expressions. 
We further extend the syntax with allocation statements \(q[n] := \talloc(d)\) for \(n\in\mathbb{N}_{>0}\), classical assignments \(x := e\), where \(x\) is a classical variable and \(e\) is a classical expression, and measurement assignments \(x := \cM[\bar{q}]\), which store the binary outcome of measuring \(\bar{q}\) with \(\cM\) in \(x\). Finally, the guards of \(\tif\) and \(\twhile\) statements are generalized from quantum measurements to classical Boolean expressions.

To accommodate classical variables, program configurations are extended from pairs \((S,\rho)\) to triples \((S,\sigma,\rho)\), where \(\sigma\in\Sigma\) is a classical state. 
Representative operational semantics rules are given below; the {\tt false} cases are analogous and omitted for brevity.
{
\[
\begin{array}{c}
\inferrule{\rho'\in\cD\pare{\cH_{\dom\rho[\sqcup/q[0],...,q[n-1]]}\otimes\cH_{q[0]}\otimes\cdots\otimes \cH_{q[n-1]}}\\
\dim\cH_{q[i]}=d\text{ for }i=0,...,n-1\\
\Tr_{q[0],...,q[n-1]}(\rho')=\rho[\sqcup/q[0],...,q[n-1]]
}{\pare{q[n]:=\talloc(d),\sigma,\rho}\to\pare{\downarrow,\sigma,\rho'}}{}\\[1em]

\inferrule{\sem{\bar{q}}^\sigma\subseteq \dom\rho}{\pare{U[\bar{q}],\sigma,\rho}\to\pare{\downarrow,\sigma,U_{\sem{\bar{q}}^\sigma}\rho U_{\sem{\bar{q}}^\sigma}^\dagger}}
\qquad
\inferrule{\;}{\pare{x:=e,\sigma,\rho}\to \pare{\downarrow,\sigma[x\mapsto\sem{e}^\sigma],\rho}}
\\
\inferrule{\sem{\bar{q}}^\sigma\subseteq \dom\rho\\ \cM=\{P_\TT,P_\FF\}}
{\pare{x:=\cM[\bar{q}],\sigma,\rho}\to
 \pare{\downarrow,\sigma[x\mapsto 1],
 P_{\TT,\sem{\bar{q}}^\sigma}\rho P_{\TT,\sem{\bar{q}}^\sigma}}}
\qquad
 \inferrule{\sem{b}^\sigma=\TT}{\pare{\tif\ b\ \tthen\ S_1\ \telse\ S_2,\sigma,\rho}\to \pare{S_1,\sigma,\rho}}{}
\end{array}
\]}\\
where \(\sem{e}^{\sigma}\) denotes the value of \(e\) under \(\sigma\), and \(\sigma[x\mapsto n]\) denotes the classical state obtained by updating \(x\) to \(n\).
Intuitively, the assignment statement \(x:=e\) updates only the classical state, whereas the measurement assignment \(x:=\cM[\bar{q}]\) updates both components: the measurement outcome is stored in \(x\), and the quantum state is projected onto the corresponding post-measurement state.
\begin{example}
Consider the program
\(
S\equiv q[2]:=\talloc(2);\ x:=\cM_{01}[q[0]];\ H[q[x]].
\)
One possible execution of this program is:
\[
\begin{aligned}
    &\pare{S,\emptyset,1}
    \to
    \pare{x:=\cM_{01}[q[0]];H[q[x]],\emptyset,|00\rangle_{q[0],q[1]}\langle 00|}\to \\
    &
    \pare{H[q[x]],x\mapsto 0,|00\rangle_{q[0],q[1]}\langle 00|}
    \to
    \pare{\downarrow,x\mapsto 0,|\!+\!0\rangle_{q[0],q[1]}\langle +0|} ,
\end{aligned}
\]
where the measurement outcome is \(0\), and consequently the operation \(H[q[x]]\) is resolved to \(H[q[0]]\).
\end{example}

\paragraph{Separation Logic with Classical Variables}
The program specifications introduced in Section~\ref{sec:separation} extend straightforwardly to executions of the form
\(
(S,\sigma,\rho)\to^*(\downarrow,\sigma',\rho').
\)
To illustrate the treatment of classical variables in the proof system, we present two representative inference rules for measurement and \(\tif\) statements.
The rule for measurement assignments is given by
{
\[
\inferrule{
\bar{q}\subseteq\bar{q}'\\
\cM=\{P_\TT,P_\FF\}\\
P\in\cP(\cH_{\bar{q}'})\\
P_1=\supp\pare{P_{\TT,\bar{q}'} P P_{\TT,\bar{q}'}}\\
P_2=\supp\pare{P_{\FF,\bar{q}'} P P_{\FF,\bar{q}'}}
}{
\triple{
\bar{q}'\mapsto P *
\Bigl(
(\bar{q}'\mapsto P_1 \sepimp \psi[1/x])
\wedge
(\bar{q}'\mapsto P_2 \sepimp \psi[0/x])
\Bigr)
}{
x:=\cM[\bar{q}]
}{
\psi
}
}.
\]
}
Intuitively, a measurement assignment can be viewed as consisting of two stages.
First, the quantum state is projected onto the post-measurement state corresponding to the observed outcome.
This stage is handled analogously to the rule for unitary transformations: the precondition guarantees that the resulting state satisfies \(\psi[1/x]\) or \(\psi[0/x]\), depending on whether the outcome is \(\TT\) or \(\FF\).
Second, the measurement outcome is recorded in the classical variable \(x\), which is reflected by the substitution of \(x\) in the corresponding postcondition: \(\sigma,\rho\models \psi[0/x]\iff \sigma[x\mapsto 0],\rho\models\psi\).

The rule for \(\tif\) statements is:
{
\[
\inferrule{
\triple{b\wedge \varphi}{S_1}{\psi}\\
\triple{\neg b\wedge \varphi}{S_2}{\psi}
}{
\triple{\varphi}{\tif\ b\ \tthen\ S_1\ \telse\ S_2}{\psi}
},
\]}
which coincides with the standard conditional rule in classical Hoare logic.
Its soundness follows directly from the fact that every state satisfying \(\varphi\) belongs to exactly one of the two cases determined by the Boolean guard \(b\):
\[
\{(\sigma,\rho)\mid \sigma,\rho\models\varphi\}
=
\{(\sigma,\rho)\mid \sigma,\rho\models\varphi,\ \sem{b}^{\sigma}=\TT\}
\;\cup\;
\{(\sigma,\rho)\mid \sigma,\rho\models\varphi,\ \sem{b}^{\sigma}=\FF\}.
\]
Moreover, by the semantics of quantum bunched logic introduced above, \(\sigma,\rho\models b\wedge \varphi\) if and only if \(\sigma,\rho\models \varphi\) and \(\sem{b}^\sigma=\TT\), and similarly for \(\neg b\wedge \varphi\).
Consequently, the precondition \(\varphi\) is partitioned according to the value of the guard, and the validity of the two premises implies the validity of the conclusion.
\begin{example}
    Consider the initialization program
    \[
    q[1]:=\talloc(2);\ x:=\cM_{01}[q[0]];\ \tif\ x=1\ \tthen\ X[q[0]]\ \telse\ \tskip.
    \]
    Let
    \(
    \psi
    =
    (x=1\wedge q[0]\mapsto |1\rangle\langle 1|)
    \vee
    (x=0\wedge q[0]\mapsto |0\rangle\langle 0|)
    \).
    By the rules for conditional statements and unitary transformations, we derive
    \[
    \triple{\psi}
    {\tif\ x=1\ \tthen\ X[q[0]]\ \telse\ \tskip}
    {q[0]\mapsto |0\rangle\langle 0|}.
    \]
    Observing that
    \(
    \psi[0/x]=q[0]\mapsto |0\rangle\langle 0|\) and \(\psi[1/x]=q[0]\mapsto |1\rangle\langle 1|\), 
    the application of the rule for measurement assignments therefore yields
    \(
    \triple{q[0]\mapsto I}
    {x:=\cM_{01}[q[0]]}
    {\psi}.
    \)
    Finally, by the allocation rule and sequential composition, we obtain
    \[
    \triple{\top^*}
    {q[1]:=\talloc(2);\ x:=\cM_{01}[q[0]];\ \tif\ x=1\ \tthen\ X[q[0]]\ \telse\ \tskip}
    {q[0]\mapsto |0\rangle\langle 0|}.
    \]
    This specification states that, starting from an empty heap, the program always terminates with \(q[0]\) in the state \(|0\>\), regardless of its initial state. Hence, the program correctly implements the intended initialization procedure.
\end{example}

\section{Case Studies}

In this section, we apply our framework to verify two practical quantum programs that involve local memory management and the allocation of qubits in unknown states, thereby demonstrating the expressiveness and effectiveness of our quantum separation logic.

\subsection{Functional Correctness and Safe Uncomputation: Program with Dirty Qubits}

\begin{figure}[t]
    \centering
    \fbox{%
    \begin{minipage}{0.94\textwidth}
        \centering
        \begin{minipage}{0.46\textwidth}
            \centering
            \hspace{2em}
            \begin{quantikz}[column sep=0.15cm, row sep=0.35cm]
                \lstick{$q_1$} &       \qw& \ctrl{1} &       \qw& \ctrl{1} & \qw&\rstick{$q_1$}\\
                \lstick{$q_2$} &       \qw& \ctrl{2} &       \qw& \ctrl{2} & \qw&\rstick{$q_2$}\\
                \lstick{$q_3$} & \ctrl{1} &       \qw& \ctrl{1} &       \qw& \qw&\rstick{$q_3$}\\
                \lstick{$a$}   & \ctrl{1} & \targ{}  & \ctrl{1} & \targ{}  & \qw&\rstick{$a$}\\
                \lstick{$q$}   & \targ{}  &       \qw& \targ{}  &       \qw& \qw&\rstick{$q+q_1q_2q_3$}
            \end{quantikz}
            \caption{Circuit implementing a 3-controlled \(X\) gate with a dirty ancilla qubit.}
            \label{fig:MCX}
        \end{minipage}
        \hfill
        \vrule
        \hfill
        \begin{minipage}{0.48\textwidth}
            \centering
            \[
            \begin{array}{ll}
                & \blueb{q_1,q_2,q_3,q\mapsto|c_1,c_2,c_3,t\>}\\
                S\equiv & a:=\talloc(2);\\
                &\hspace{1em}\begin{cases}
                    {\color{purple}\left \{q_1,q_2,q_3,q\mapsto I* a,a'\mapsto|\Phi\>\right \}}\\
                    \mathrm{Toffoli}(q_3,a,q);\mathrm{Toffoli}(q_1,q_2,a);\\
                    \mathrm{Toffoli}(q_3,a,q);\mathrm{Toffoli}(q_1,q_2,a);\\
                    {\color{purple}\left \{ a,a'\hookrightarrow |\Phi\>\right \}}
                 \end{cases}\\
                & \trelease(a);\\
                & \blueb{q_1,q_2,q_3,q\mapsto|c_1,c_2,c_3,t+c_1c_2c_3\>}
            \end{array}
            \]
            \vspace{-1em}
            \caption{Program for the 3-controlled \(X\) gate.}
            \label{fig:mcx-implement}
            \begin{tikzpicture}[remember picture, overlay]
                \draw (-2.5, 2.9) node[font=\small] {safely};
                \draw (-2.5, 2.5) node[font=\small] {uncompute};
            \end{tikzpicture}
        \end{minipage}
    \end{minipage}
    }
\end{figure}
We first consider a program implementing a 3-controlled \(X\) gate using four Toffoli gates and one dirty ancilla qubit \(a\)~\cite{gidneyblog}, as shown in Figure~\ref{fig:MCX}.
Operationally, the circuit realizes the classical transformation
\(
(c_1,c_2,c_3,t)\mapsto (c_1,c_2,c_3,t+c_1c_2c_3)
\) coherently on computational-basis states, for \(c_1,c_2,c_3,t\in\bZ_2\), stored in qubits \(q_1,q_2,q_3\) and \(q\), while restoring the ancilla qubit \(a\) to its original state. 
Importantly, the initial state of \(a\) is arbitrary, including being entangled with other qubits.

Two properties are expected from this program, both of which admit local specifications in our framework:
\begin{enumerate}
    \item \emph{Correctness}: the program implements the intended transformation, which can be specified by the blue triple shown in Figure~\ref{fig:mcx-implement}:
    \[
    \hspace{3em}
    \triple
    {q_1,q_2,q_3,q\mapsto|c_1,c_2,c_3,t\>}
    {S}
    {q_1,q_2,q_3,q\mapsto|c_1,c_2,c_3,t+c_1c_2c_3\>},
    \]
    where we write \(\bar q\mapsto|C\>\) as shorthand for
    \(\bar q\mapsto|C\>\<C|\).
    The specification is local in the sense that the pre- and postconditions describe precisely the heap fragment consisting of the qubits \(q_1,q_2,q_3,q\) involved in the transformation, while arbitrary surrounding heap resources can be incorporated via separating conjunction.

    \item \emph{Safety}: the program safely uncomputes the dirty ancilla qubit \(a\), restoring it to its original state unconditionally.
    By Theorem~6.1 of~\cite{dirtyqubit}, this property can be specified by the red triple shown in Figure~\ref{fig:mcx-implement}:
    \[
    \hspace{3em}
    \purpleb{
    q_1,q_2,q_3,q\mapsto I
    \;*\;
    a,a'\mapsto|\Phi\>}
    \;
    \mathrm{Toffoli}(q_3,a,q);
    \cdots;
    \mathrm{Toffoli}(q_1,q_2,a)
    \;
    \purpleb{
    a,a'\hookrightarrow|\Phi\>},
    \]
    where
    \(
    |\Phi\>
    =
    \frac{1}{\sqrt 2}(|00\>+|11\>)
    \)
    is a maximally entangled state.
    This specification is likewise local: the precondition contains exactly the heap resources required by the program body, including the control and target qubits \(q_1,q_2,q_3,q\), which must be explicitly specified to guarantee memory-safe execution as discussed in Section~\ref{sec:separation}. The auxiliary qubit \(a'\) is introduced as a hypothetical reference system initially maximally entangled with \(a\), and the postcondition asserts that this entanglement is preserved, thereby establishing that the overall effect on \(a\) is the identity~\cite{dirtyqubit}.
\end{enumerate}
Both specifications are derivable in the proof system of Section~\ref{sec:separation}, thereby establishing the desired correctness and safety properties. 

\subsection{Loop with Local Variables: Repeat-Until-Success Circuit}\label{sec:cases-rus}
\begin{figure}[t]
    \centering
    \fbox{%
    \begin{minipage}{0.94\textwidth}
        \centering
        \begin{minipage}{0.46\textwidth}
            \centering
            \begin{quantikz}[column sep=0.25cm, row sep=0.4cm]
                \lstick{$q_1:|+\>$}       & \ctrl{1} &\qw       &\qw       &\qw       & \ctrl{1}\gategroup[2,steps=1,style={inner sep=0pt,dashed, inner xsep=0pt, inner ysep=0pt},background]{}      & \meter{}\\
                \lstick{$q_2:|+\>$}       & \ctrl{1} &\qw       &\qw       &\qw       & \gate{Z}                                                       & \meter{}\\
                \lstick{$q_3:|0\>$}       & \targ{}  & \ctrl{1} &\qw       & \ctrl{1} & \meter{} \wire[u][1]{c}\\
                \lstick{$q:|\lambda\>$}    &\qw       & \targ{}  & \gate{S} & \targ{}  & \gate{Z} & \rstick{$V_3|\lambda\>$}
            \end{quantikz}
            \caption{The repeat-until-success circuit for implementing a single-qubit unitary \(V_3\) using ancilla qubits, where the operation enclosed by the dashed box is classically controlled by the measurement outcome.
            Adapted from Figure~1(b) of~\cite{rus}.
            }
            \label{fig:rus}
        \end{minipage}
        \hfill
        \vrule
        \hfill
        \hspace{-1.3em}
        \begin{minipage}{0.49\textwidth}
            \centering
            \small
            \[
            \hspace{-1em}
             \begin{array}{ll}
                & \blueb{q\mapsto|\lambda\>}\,q_1:=\talloc(2);q_2:=\talloc(2);\\
                &[q_1,q_2]:=|0\>;X[q_1];H[q_1];\\
                & \blueb{q_1,q_2\mapsto|\!-\!0\>*q\mapsto|\lambda\>}\implies\blue{\{(q_1,q_2\hookrightarrow I)\wedge}\\
                &\blue{(q_1,q_2\hookrightarrow P_\TT\sasaki \varphi)\wedge (q_1,q_2\hookrightarrow P_\FF\sasaki\psi)\}}\\
                & \twhile\ \cM_{1}[q_1,q_2]\ \tdo\\
                & \hspace{3.0em} \blueb{\varphi:= q_1,q_2\mapsto I*q\mapsto|\lambda\>\<\lambda|}\\
                & \hspace{3.0em}q_3:=\talloc(2);[q_1,q_2,q_3]:=|0\>;\\
                & \hspace{3.0em}H[q_1];H[q_2];\mathrm{Toffoli}[q_1,q_2,q_3];\\
                & \hspace{3.0em}\mathrm{CNOT}[q_3,q];S[q];\mathrm{CNOT}[q_3,q];Z[q];\\
                & \hspace{3.0em}\tif\ \cM_{2}[q_3]\ \textbf{then}\ \mathrm{CZ}[q_1,q_2];\ \textbf{else}\ \tskip;\\
                & \hspace{3.0em}\trelease(q_3)\blue{\{(q_1,q_2\hookrightarrow I)\wedge}\\
                & \hspace{3.0em}\blue{(q_1,q_2\!\hookrightarrow\!P_\TT\!\sasaki\!\varphi)\!\wedge\!(q_1,q_2\!\hookrightarrow\!P_\FF\!\sasaki\!\psi)\}}\\
                &\tend;\blueb{\psi:= q_1,q_2\mapsto I*q\mapsto V_3|\lambda\>\<\lambda|V_3^\dagger}\\
                &\trelease(q_2);\trelease(q_1)\blueb{q\mapsto V_3|\lambda\>}
            \end{array}
            \]
            \vspace{-1em}
            \caption{The program implementing the repeat-until-success circuit, along with an inline correctness specification.}
            \label{fig:rus-implement}
        \end{minipage}
    \end{minipage}%
    }
\end{figure}

We next present a case study that illustrates reasoning about a quantum program containing a while-loop and local variables whose scope is confined to the loop body.
The program employs a \emph{repeat-until-success} (RUS) strategy~\cite{rus} to implement the single-qubit unitary
\(
V_3=\pare{I+2iZ}/{\sqrt{5}},
\)
as shown in Figure~\ref{fig:rus}. 
Operationally, the circuit is repeatedly executed until both ancillary qubits are measured in the \(+1\) eigenstate, at which point the desired \(V_3\) transformation is obtained.
The corresponding program is shown in Figure~\ref{fig:rus-implement}, where the two measurements are defined by
\[
\cM_1=
\bigl\{
P_\TT = I-|\!++\rangle\langle ++\!|,
\;
P_\FF = |\!++\rangle\langle ++\!|
\bigr\},
\qquad
\cM_2=
\bigl\{
P_\TT = |-\rangle\langle -|,
\;
P_\FF = |+\rangle\langle +|
\bigr\}.
\]

The correctness specification, restricted to the relevant local piece of the heap, is given by
\[
\triple{q\mapsto |\lambda\>\<\lambda|}
       {q_1:=\talloc(2);\ldots;\trelease(q_1)}
       {q\mapsto V_3|\lambda\>\<\lambda|V_3^\dagger},
\]
where \(|\lambda\rangle\) is an arbitrary state of the target qubit \(q\).
An inlined proof outline is shown in Figure~\ref{fig:rus-implement}, illustrating how loop reasoning is carried out using a loop invariant.
We choose the loop invariant
\(
\varphi := q_1,q_2\mapsto I * q\mapsto |\lambda\>\<\lambda|,
\)
and the desired postcondition upon loop termination
\(
\psi := q_1,q_2\mapsto I * q\mapsto V_3|\lambda\>\<\lambda|V_3^\dagger.
\)
Applying the inference rule for while-loops in Figure~\ref{fig:qsep-proof-system} reduces the verification of the entire loop to a single verification condition for the loop body:
\[
\triple{\varphi}
       {q_3:=\talloc(2);\ldots;\trelease(q_3)}
       {q_1,q_2\hookrightarrow I
        \wedge
        q_1,q_2\hookrightarrow P_\FF \sasaki \psi
        \wedge
        q_1,q_2\hookrightarrow P_\TT \sasaki \varphi}.
\]
Intuitively, this condition states that the loop body either establishes the target postcondition when the measurement outcome is \(P_\FF\), or restores the invariant when the outcome is \(P_\TT\).
The verification can then be completed by a static, sequential reasoning process over the loop body, without explicitly unrolling or repeatedly executing the loop.
The detailed computation is deferred to the appendix.

\section{Related Work}\label{sec:related}

\paragraph{Separation Logic} \citet{bias} and \citet{reynolds} pioneered the use of bunched logic as an assertion language for reasoning about heap resources. 
Various extensions of separation logic have been explored across different resource models, including permissions \cite{SL-permission}, concurrency \cite{SL-concurrent}, time and auxiliary state \cite{SL-statetime}, and protocols \cite{SL-protocol}. 

\paragraph{Bunched Logic} \citet{logicofbi} introduced bunched logic as a substructural logic suitable for describing resource composition. 
Bunched logic has found applications in various contexts such as type theory \cite{bi-type1, bi-type2}, game theory \cite{bi-games}, and quantum computation \cite{qsep}. It has also been extended to characterize probabilistic \cite{bi-quantative} and relational properties \cite{bi-relational} of programs with heap manipulations.

\paragraph{Quantum Logic} Two main approaches based on different interpretations of quantum assertions have emerged. The first approach, proposed by Birkhoff and von Neumann~\cite{quantumlogic}, associates quantum events with projection operators which form an orthomodular lattice. 
This approach has been further developed using categorical and algebraic methods \cite{qlogic1, qlogic2, qlogic3}, and reformulated as orthologic, whose proof theory and algorithms have been well-studied \cite{orthologic1, orthologic2, orthologic3}. The second approach associates quantum events with physical observables \cite{qhoare}, which generalize projection operators to describe probabilistic properties and retain an algebraic structure known as an effect algebra \cite{effectalgebra1, effectalgebra2, effectalgebra3}.

\paragraph{Formal Verification of Quantum Programs} As reviewed in~\cite{CVLreview}, software engineering methods have been adapted for quantum computing, including model checking \cite{modelchecking}, testing and debugging \cite{test1, test2}, abstract interpretation \cite{qabstractinterpretation,FENG2023105077}, ZX calculus for quantum circuits~\cite{coecke2011interacting, PyZX} and various automatic/interactive verifiers \cite{qafny,automated, VyZX23, chareton2020deductive}. 
Predicate transformers for quantum computation were introduced by~\citet{qwp}. 
Building on this foundation, \citet{qhoare} developed the first sound and relatively complete quantum Hoare logic from an extensional perspective. More recently, \citet{expressive} proposed an expressive assertion language and established intensional completeness.
In addition, quantum Hoare logic has been further refined and adapted for a variety of practical purposes~\cite{LSZreview}, including quantum error-correcting programs~\cite{vqec}, the incorporation of projection-based assertions~\cite{aQHL,ghost,relational1}, support for classical variables~\cite{cqhl}, and the development of quantum separation logics~\cite{qsep,qsep1,qsep2,rapunsl}.

\bibliographystyle{ACM-Reference-Format}
\bibliography{refs}


\end{document}